\documentclass[aps,prl,reprint,superscriptaddress]{revtex4-1}

\usepackage{amsmath,amssymb}
\usepackage{braket}
\usepackage{graphicx}
\usepackage{subfig}
\usepackage{bm}

\begin{document}
\title{Efficient simulation of quantum error correction under coherent error based on non-unitary free-fermionic formalism}
\author{Yasunari Suzuki}
\email[]{suzuki@qi.t.u-tokyo.ac.jp}
\affiliation{Department of Applied Physics, Graduate School of Engineering, The University of Tokyo, 7-3-1 Hongo, Bunkyo-ku, Tokyo 113-8656, Japan}
\affiliation{Photon Science Center, Graduate School of Engineering, The University of Tokyo, 7-3-1 Hongo, Bunkyo-ku, Tokyo 113-8656, Japan}
\author{Keisuke Fujii}
\email[]{fujii@qi.t.u-tokyo.ac.jp}
\affiliation{Department of Applied Physics, Graduate School of Engineering, The University of Tokyo, 7-3-1 Hongo, Bunkyo-ku, Tokyo 113-8656, Japan}
\affiliation{Photon Science Center, Graduate School of Engineering, The University of Tokyo, 7-3-1 Hongo, Bunkyo-ku, Tokyo 113-8656, Japan}
\affiliation{JST, PRESTO, 4-1-8 Honcho, Kawaguchi, Saitama, 332-0012, Japan}
\author{Masato Koashi}
\email[]{koashi@qi.t.u-tokyo.ac.jp}
\affiliation{Department of Applied Physics, Graduate School of Engineering, The University of Tokyo, 7-3-1 Hongo, Bunkyo-ku, Tokyo 113-8656, Japan}
\affiliation{Photon Science Center, Graduate School of Engineering, The University of Tokyo, 7-3-1 Hongo, Bunkyo-ku, Tokyo 113-8656, Japan}
\date{\today}

\begin{abstract}
In order to realize fault-tolerant quantum computation, tight evaluation of error threshold under practical noise models is essential. 
While non-Clifford noise is ubiquitous in experiments, the error threshold under non-Clifford noise cannot be efficiently treated with known approaches. 
We construct an efficient scheme for estimating the error threshold of one-dimensional quantum repetition code under non-Clifford noise.
To this end, we employ non-unitary free-fermionic formalism for efficient simulation of the one-dimensional repetition code under coherent noise. 
This allows us to evaluate the effect of coherence in noise on the error threshold without any approximation.
The result shows that the error threshold becomes one third when noise is fully coherent. 
Our scheme is also applicable to the surface code undergoing a specific coherent noise model.
The dependence of the error threshold on noise coherence can be explained with a leading-order analysis with respect to coherence terms in the noise map. 
We expect that this analysis is also valid for the surface code since it is a two-dimensional extension of the one-dimensional repetition code. 
Moreover, since the obtained threshold is accurate, our results can be used as a benchmark for approximation or heuristic schemes for non-Clifford noise.
\end{abstract}

\maketitle

{\it Introduction.---}
Quantum error correction (QEC) is a key technology for building a scalable fault-tolerant quantum computer. According to the theory of fault-tolerant quantum computation, one can perform quantum computation with arbitrary accuracy if the error probability is below a certain threshold value \cite{kitaev1997quantum,knill1998resilient,aharonov1997fault}. The threshold values of various QEC schemes have been calculated under various assumptions of the noise models and degrees of rigor \cite{fern2004generalized,greenbaum2016modeling,wang2003confinement,wang2011surface,fowler2012towards,stephens2014fault,ghosh2012surface,geller2013efficient,gutierrez2013approximation,gutierrez2015comparison,magesan2013modeling,puzzuoli2014tractable,gutierrez2016errors,rahn2002exact,chamberland2016hard,tomita2014low,ferris2014tensor,darmawan2016tensor}.
In the case of the noise model which only consists of probabilistic Clifford gates and Pauli measurement channels, such as depolarizing noise, the threshold value can be efficiently and accurately estimated numerically \cite{wang2003confinement,wang2011surface,fowler2012towards,stephens2014fault} by virtue of the Gottesman-Knill theorem \cite{gottesman1998heisenberg,aaronson2004improved}. 
On the other hand, non-Clifford noise is unavoidable in practical experiments \cite{kelly2015state,corcoles2015demonstration,riste2015detecting}, but QEC circuits under non-Clifford noise cannot be treated with this approach. Specifically, it is theoretically predicted that coherent noise, which is non-Clifford and is caused, for example, by over rotation, can have negative effects on quantum error correction \cite{kueng2015comparing}. Therefore, massive effort has been made for evaluating the effect of noise coherence on the error threshold. 
Since the simulation of quantum circuits under arbitrary local noise sometimes becomes as hard as that of universal quantum computation, we cannot efficiently simulate QEC circuits under coherent noise with straightforward methods. While its computational cost can be relaxed in some extent \cite{tomita2014low,ferris2014tensor,darmawan2016tensor}, the tractable number of qubits with straightforward methods is limited. In the case of concatenated codes, there is an efficient method to analytically estimate the error threshold under non-Clifford noise \cite{rahn2002exact}. However, this technique is not applicable to topological codes, which are more feasible in practical experiments \cite{kelly2015state,corcoles2015demonstration,riste2015detecting,lidar2013quantum}. In general, we may approximate non-Clifford noise by a Clifford channel for an efficient simulation  \cite{gutierrez2016errors,ghosh2012surface,geller2013efficient,gutierrez2013approximation,gutierrez2015comparison,magesan2013modeling,puzzuoli2014tractable}, but the accuracy of the estimated threshold is sacrificed. An efficient and accurate scheme, which works for topological codes under non-Clifford noise, is still lacking.

Here, we construct an efficient and accurate scheme to simulate one-dimensional (1D) repetition code with repetitive parity measurements under coherent noise. While the 1D repetition code cannot protect a logical qubit from arbitrary single-qubit error, it is still able to capture a necessary ingredient for fault-tolerant QEC, and hence was experimentally demonstrated as a building block for scalable fault-tolerant quantum computation \cite{kelly2015state,riste2015detecting}. The key idea in our scheme is reducing the QEC circuit of the 1D repetition code to a classically simulatable class of non-unitary free-fermionic dynamics, which is known as a variation of matchgate quantum computing \cite{valiant2002quantum,terhal2002classical,knill2001fermionic,bravyi2004lagrangian,bravyi2005classical,jozsa2008matchgates,jozsa2013jordan,brod2016efficient,bravyi2014efficient}.
As compared to the stochastic noise model, we find that the error threshold of the 1D repetition code becomes about one third when noise is fully coherent. The dependence of the error threshold on noise coherence is explained by using a leading-order analysis with respect to coherence terms of noise map. We expect that a similar analysis holds in the surface code \cite{dennis2002topological,bravyi1998quantum,fowler2012surface}, which is the most experimentally feasible QEC scheme, since the surface code is a two-dimensional extension of the 1D repetition code.
\begin{figure}[tp]
 \centering
 \includegraphics[clip, width=7.5cm]{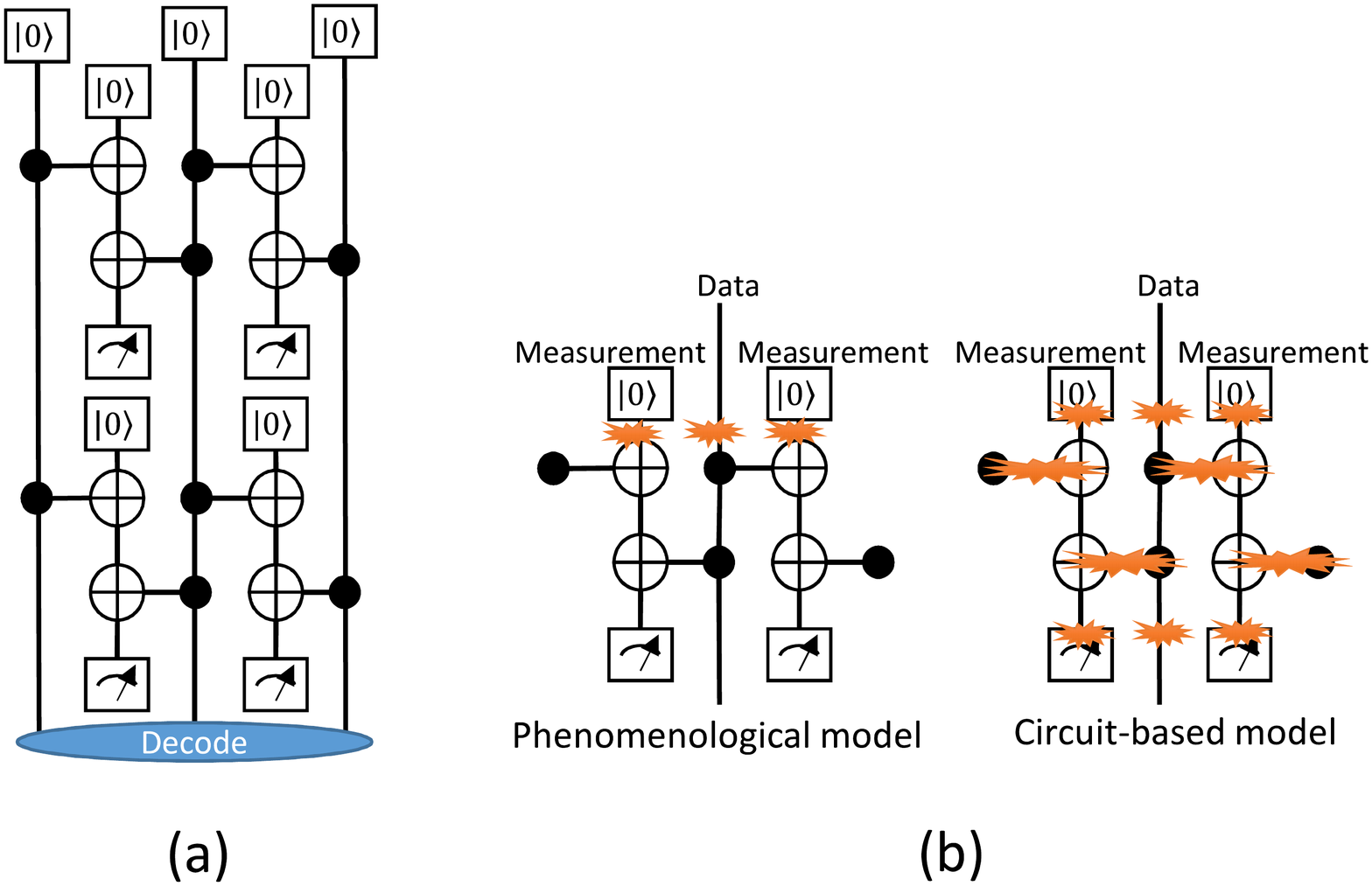}
 \caption{(a) The QEC circuit of the 1D repetition code with $n=3$.  (b) The error allocation of the phenomenological (left) and circuit-based (right) models.}\label{fig:1all}
\end{figure}
Furthermore, our accurate results can be used as a benchmark for approximation or heuristic schemes for estimating the error threshold under non-Clifford noise.

{\it Simulation of the 1D repetition code.---}
The quantum circuit of the 1D repetition code with repetitive parity measurements is shown in Fig.\,\ref{fig:1all}(a). In the 1D repetition code, one logical bit $a_{\rm in}$ is encoded into physical $n$ data qubits $\{\ket{0}^{\otimes n}, \ket{1}^{\otimes n}\}$, which are stabilized by operators $\{Z_iZ_{i+1}\}_{i=1}^{n-1}$, where $A_i (A\in \{X,Y,Z\})$ is the Pauli operator on the $i$-th data qubit. Error syndrome is measured through $(n-1)$ measurement qubits, each of which monitors the parities of the neighboring data qubits, i.e., $Z_i Z_{i+1}$. 
The measurements are repetitively performed for $T$ cycles. The encoded bit is finally decoded from the data qubits and $T(n-1)$ syndromes, which can be efficiently done using minimum-weight perfect matching \cite{kelly2015state}. 
The probability with which the decoded bit $a_{\rm out}$ is flipped is defined as logical error probability $p_{\rm L}$, which is the failure probability of the decoding.

Since the 1D repetition code is capable of correcting only $X$-type error, we consider a CPTP (completely positive trace-preserving) map of a general single-qubit $X$-type noise, which is regarded as a mixture of the $X$-type unitary (fully-coherent) and stochastic (incoherent) noise: 
\begin{eqnarray}
{\mathcal E}(\rho,X) &=& ce^{i\theta X}\rho e^{-i\theta X} + (1-c)\left((1-p)\rho + pX \rho X\right) \label{eqnoise} \nonumber \\
&=& \frac{1+c}{2} e^{i\theta X} \rho e^{-i\theta X} + \frac{1-c}{2} e^{-i\theta X} \rho e^{i\theta X} {\rm ,}
\end{eqnarray}
where $\theta$ is defined by $\cos \theta  = \sqrt{1-p}$ and $\sin \theta = \sqrt{p}$. The parameter $c$ ($0\leq c \leq 1$), which we call noise coherence, is a measure of coherence in the noise. We call the parameter $p$ ($0\leq p \leq 1$) as the physical error probability since it can be understood as the probability with which the input state $\ket{0}$ is measured as the output state $\ket{1}$.
We consider two types of noise allocation models \cite{landahl2011fault} as shown in Fig.\,\ref{fig:1all}(b). In the case of the phenomenological model, a noise map ${\mathcal E(\rho,X)}$ is located on each of the data and measurement qubits at the beginning of each cycle.
In the case of the circuit-based model, the noise map is located at each time step of preparation, gate operation, and measurement, on every qubit including the one that is idle at the time step. There, we assume that two-qubit noise map $p_{XI}{\mathcal E}(\rho, X \otimes I)+ p_{IX}{\mathcal E}(\rho, I \otimes X)+p_{XX}{\mathcal E}(\rho, X \otimes X)$ acts on the output qubits after each controlled-Not (CNOT) operation. Since our noise models are symmetric over the bit values, we may choose $a_{\rm in} = 0$ to evaluate the logical error probability as $p_{\rm L} = {\rm Pr}(a_{\rm out} = 1)$.

Before reducing the noisy circuit to free-fermionic dynamics, we reformulate it as a sequence of generalized measurements on the data qubits, such that the state after each measurement is pure. 
We denote the outcome of the $k$-th measurement by $t_k$ and the corresponding Kraus operator by $K_k^{(t_k)}$. The probability of a sequence of outcomes ${\bm t}_k := t_k ... t_1$ is given by
\begin{eqnarray}
{\rm Pr}({\bm t}_{k}) = \Gamma({\bm t}_k) := \braket{0^{\otimes n} | ({\bm K}^{({\bm t}_k)})^{\dagger} {\bm K}^{({\bm t}_k)} | 0^{\otimes n}} \label{eqprob} {\rm ,}
\end{eqnarray}
where ${\bm K}^{({\bm t}_k)} := K_k^{(t_k)} K_{k-1}^{(t_{k-1})} ... K_1^{(t_1)}$.
We may identify three types of operations on data qubits to assign Kraus operators. For clarity, we describe the case of the phenomenological model.
The first type is the single-qubit noise ${\mathcal E}$ given in Eq.(\ref{eqnoise}). Its operation on the $i$-th qubits is equivalently described by Kraus operators $K_{{\rm noise},i}^{(\phi)} = \sqrt{p(\phi)} e^{i\phi X_i}$, where $\phi \in \{\pm \theta \}$ and $p(\pm \theta) := \frac{1\pm c}{2}$. 
The second type is the parity measurement on the $i$-th and $(i+1)$-th data qubits, which composed of a measurement qubit and two CNOT gates. Treating the noise map $\mathcal E$ on the measurement qubit as above, it can be represented by $K^{(s,\phi)}_{{\rm parity},i} = \sqrt{p(\phi)} \frac{1}{2}(I + (-1)^{s} e^{-2i \phi} Z_iZ_{i+1})$, where $s \in \{0,1\}$ is the output of the parity measurement. In the case of the circuit-based model, we may still use the same form of $K^{(s,\phi)}_{{\rm parity},i}$ except for varying the probability mass function $p(\phi)$ (see Appendix A \cite{SM}).
The third type appears in an alternate description of the decoding process. Though the input bit is usually decoded through noisy direct measurements of the data qubits and classical computation, we use the following equivalent process instead. We apply map $\mathcal E$ on each data qubit. We perform ideal parity measurements on neighboring data qubits, whose Kraus operator is given by $K^{(s)}_{{\rm parity},i} = \frac{1}{2}(I + (-1)^{s} Z_iZ_{i+1})$. 
Let $k_f$ be the index of the last ideal parity measurement, and define ${\bm t} := {\bm t}_{k_f}$. 
Based on all the measured parities, which is included in $\bm t$, we choose a recovery operation $R({\bm t})=\prod_{i=1}^n X_i^{r_i({\bm t})}$, where $r_i({\bm t}) \in \{0,1\}$ is determined using minimum-weight perfect matching \cite{kelly2015state,edmonds1965paths,kolmogorov2009blossom}. 
The recovered state $R({\bm t}){\bm K}_{k_f}^{({\bm t})} \ket{0}^{\otimes n}$ is in the code space of the 1D repetition code, which should be written in the form $\alpha \ket{0}^{\otimes n} + \beta \ket{1}^{\otimes n}$. The decoded bit $a_{\rm out}$ is thus obtained by measuring the $n$-th qubit. The joint probability of obtaining $\bm t$ and a decoding failure is then given by
\begin{eqnarray}
 {\rm Pr}(a_{\rm out}&=&1, {\bm t}) = \Gamma_{\rm L}({\bm t}) \nonumber \\
&{}& := \braket{0^{\otimes n}|(R({\bm t}) {\bm K}^{({\bm t})})^{\dagger} \frac{I-Z_n}{2} R({\bm t}) {\bm K}^{({\bm t})}| 0^{\otimes n}}. \label{eqpl}
\end{eqnarray}
From Eqs. (\ref{eqprob}) and (\ref{eqpl}), we have,
\begin{eqnarray}
p_{\rm L} = {\rm Pr}(a_{\rm out} = 1) = \braket{\Gamma_{\rm L}({\bm t}) / \Gamma({\bm t})}_{\bm t} ,
\end{eqnarray}
which means that we can accurately calculate $p_{\rm L}$ by sampling $\bm t$ with probability $\Gamma(\bm t)$ repeatedly and by taking the average of $\Gamma_{\rm L}({\bm t})/\Gamma({\bm t})$. Since the sampling of $\bm t$ can be done by sequentially generating $t_k$ according to ${\rm Pr}(t_k | {\bm t}_{k-1}) = \Gamma({\bm t}_k) / \Gamma({\bm t}_{k-1})$, the efficiency of this scheme follows that of computing $\Gamma({\bm t}_k)$ and $\Gamma_{\rm L}(\bm t)$.

{\it Reduction to non-unitary free-fermionic dynamics.---}
We use non-unitary free-fermionic dynamics to calculate $\Gamma({\bm t}_k)$ and $\Gamma_{\rm L}({\bm t})$ efficiently. Let us briefly summarize the known facts about non-unitary free-fermionic dynamics \cite{bravyi2004lagrangian,bravyi2014efficient}.
We define $\{c_i\}$ ($1 \leq i \leq 2n$) as the Majorana fermionic operators for $n$ fermionic modes, which satisfy $\{c_i, c_j\} = 2\delta_{i,j}$, $c_i^{\dagger} = c_i$ and $c_i^2 = I$. 
The covariance matrix for a pure state $\ket{\psi}$ is defined as $M_{ij}=\frac{-i}{\braket{\psi|\psi}}\frac{\braket{\psi |[c_i,c_j]|\psi}}{2}$.
We call the state $\ket{\psi}$ is a fermionic Gaussian state (FGS) iff the covariance matrix satisfies $MM^T=I$. An FGS can be fully specified by a pair $(M,\Gamma)$, where $\Gamma$ is the norm $\braket{\psi|\psi}$. The absolute value of the inner product of two FGSs, $\ket{\psi} \mapsto (M_{\psi},\Gamma_{\psi})$ and $\ket{\phi} \mapsto (M_{\phi},\Gamma_{\phi})$, is given by $|\braket{\psi|\phi}|^2 = 2^{-n}\Gamma_{\psi} \Gamma_{\phi} \det(M_{\psi}+M_{\phi})$.
An operator of form $e^{\sum_{i<j} \alpha_{ij} c_i c_j}$ with $\alpha_{ij}$ being a complex value is called a fermionic Gaussian operator (FGO). Note that FGOs are not necessarily unitary. An FGO maps any FGS to another FGS. Given an FGO $G$ and an input FGS $\ket{\psi} \mapsto (M,\Gamma)$, the description $(M',\Gamma')$ for the output state $G\ket{\psi}$ is calculated as follows. Consider a fermionic maximally entangled state $\ket{\psi_{\rm M}} \mapsto (M_{\rm M}, \Gamma_{\rm M})$ of $2n$ fermionic modes, which is defined by $M_{\rm M} = \left( \begin{array}{cc} 0 & I \\ -I & 0 \end{array} \right)$ and $\Gamma_{\rm M} = 1$. We calculate $(M_{\rm G},\Gamma_{\rm G})$ corresponding to the state $\ket{\psi_{\rm G}} = (G\otimes I ) \ket{\psi_{\rm M}}$. In terms of matrices $A,B,D$, which are defined by $\left( \begin{array}{cc} A& B \\ -B^T & D \end{array}\right) = M_{\rm G}$, the output is calculated as $(M',\Gamma') = (A-B(M-D)^{-1} B^T , \Gamma_{\rm G} \Gamma \sqrt{\det(M-D)})$. In this way, free-fermionic dynamics consisting of FGOs on FGSs can be simulated efficiently. 

Now we are ready to reformulate the QEC process with non-unitary free-fermionic dynamics. Using the Jordan-Wigner transformation, we may choose $c_{2i-1} = \left(\prod_{j=1}^{i-1} X_j\right) Z_i$ and $c_{2i} = \left(\prod_{j=1}^{i-1} X_j\right) Y_i$. We see that $X_i = -ic_{2i}c_{2i-1}$ and $Z_iZ_{i+1} = -i c_{2i+1}c_{2i}$, which are both quadratic terms of the Majorana fermionic operators. Therefore, $K_{{\rm noise},i}^{(\phi)}, K_{{\rm parity},i}^{(s,\phi)},K_{{\rm parity},i}^{(s)}$ and $R$ are FGOs. Unfortunately, the initial state $\ket{0}^{\otimes n}$ and the operator $\frac{I-Z_n}{2}$ for calculating $\Gamma_{\rm L}({\bm t})$ are not an FGS and an FGO, respectively. For an efficient simulation, we need a further trick as follows. (While similar tricks in Refs. \cite{bravyi2005classical,jozsa2013jordan,brod2016efficient} might be employed, the following construction is much simpler and more efficient for our purpose.) We add the $(n+1)$-th ancillary qubit and corresponding Majorana fermionic operators $c_{2n+1},c_{2n+2}$. Using an FGS $\ket{\tilde{\psi}} := (\ket{0}^{\otimes (n+1)} + \ket{1}^{\otimes (n+1)})/\sqrt{2}$, it is not difficult to show that 
\begin{eqnarray}
&{}& \Gamma({\bm t}_k) = \braket{\tilde{\psi} | ({\bm K}^{({\bm t}_k)})^{\dagger} {\bm K}^{({\bm t}_k)} | \tilde{\psi}} \label{eqpap},\\
&{}& \Gamma_{\rm L}({\bm t}) = \braket{\tilde{\psi}| (R({\bm t}) {\bm K}^{({\bm t})})^{\dagger} \frac{I-Z_nZ_{n+1}}{2} R({\bm t}) {\bm K}^{({\bm t})}| \tilde{\psi}} \label{eqplap},
\end{eqnarray}
since $K^{({\bm t}_k)}$ and $R({\bm t})$ commute with $\prod_{j=1}^{n+1} X_j$. Hence, they can be efficiently calculated (see Appendix A \cite{SM} for detail).
%

\begin{figure}[tp]
 \centering
 \includegraphics[clip, width=7.5cm]{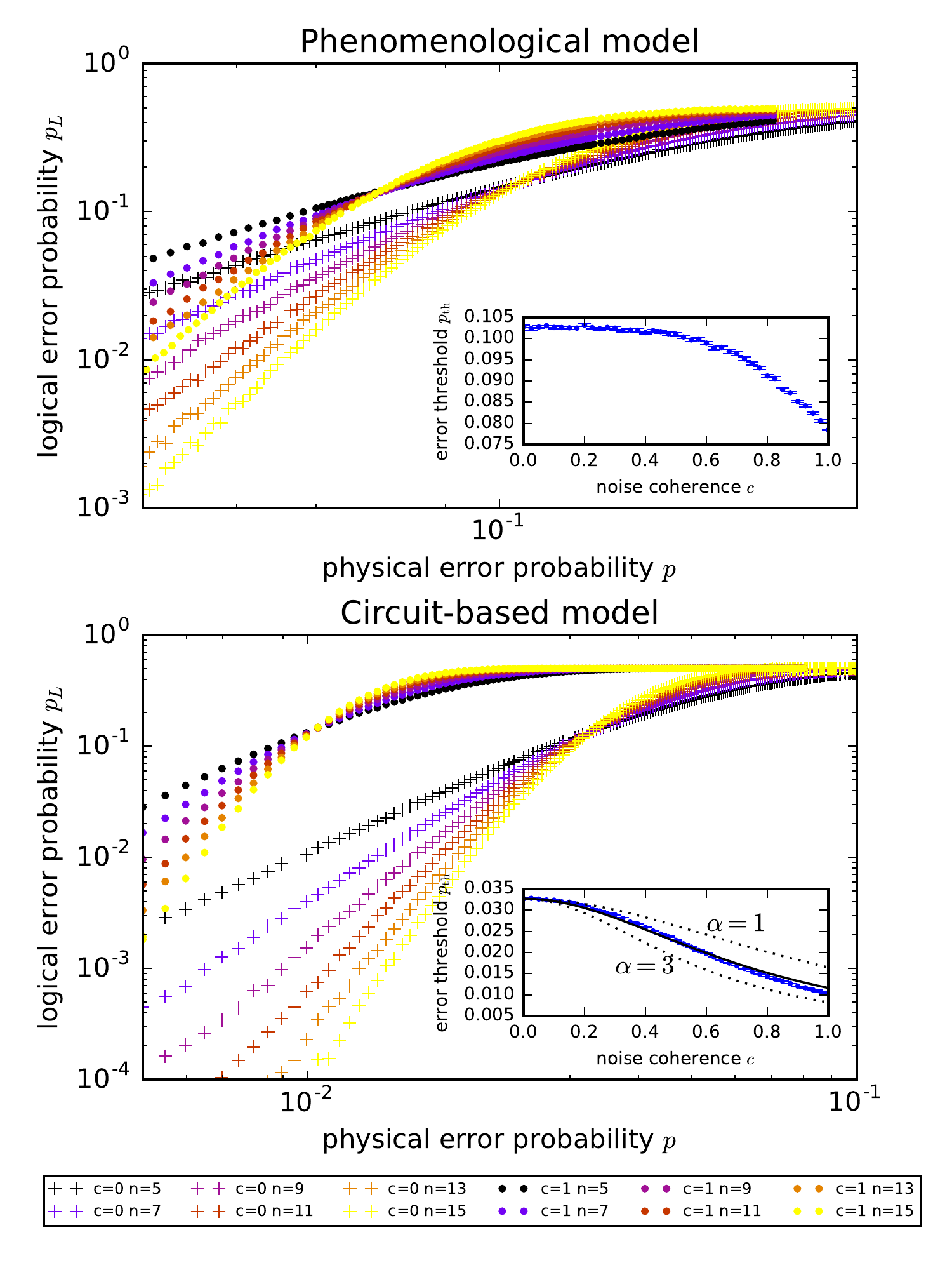}
 \caption{The logical error probability $p_{\rm L}$ is plotted as a function of the physical error probability $p$ for the two patterns of noise allocation. Insets show the error threshold $p_{\rm th}$ as a function of the amount of the coherence $c$. The blue dots are numerical results. The solid black curve in the circuit-based model is estimated behavior from the simulation of small-size QEC circuits. The dotted curves are drawn as references.}
 \label{fig:graph}
\end{figure}

{\it Result.---}
We show the logical error probability $p_{\rm L}$ as a function of the physical error probability $p$ under incoherent noise ($c=0$) and fully coherent noise ($c=1$) in Fig.\,\ref{fig:graph}. To observe clear behavior of the error threshold, we have varied the number $T$ of cycles according to $n$ as $T=n-1$. We also assumed uniform error probability for two-qubit noise, i.e. $p_{XI}=p_{IX}=p_{XX}=\frac{1}{3}$. We employed uniform weighting for performing minimum-weight perfect matching. The logical error probability $p_{\rm L}$ is expected to be exponentially small in the number of the data qubits $n$ as far as the physical error probability $p$ is below a certain value, which we call the error threshold $p_{\rm th}$. By using the scaling ansatz \cite{wang2003confinement,stephens2014fault}, we obtained the threshold values $p_{\rm th}=10.34(1)\%$ for $c=0$ and $7.87(2)\%$ for $c=1$ in the phenomenological model, and $3.243(6)\%$ for $c=0$ and $1.040(5)\%$ for $c=1$ in the circuit-based model. Our result for $c=0$ in the case of the phenomenological model is consistent with the known results \cite{wang2003confinement}. For more detailed procedures, see Appendix B \cite{SM}. 
We also confirmed exponential decay of logical error probability $p_{\rm L}$ with code distance $d$ below the threshold value, which is approximated by $p_{\rm L} \propto \left( \frac{p}{p_{\rm th}}\right)^{d/2}$ regardless of the coherence of the noises (see Appendix E \cite{SM}).

\begin{figure}[tp]
 \centering
 \includegraphics[clip ,width=5cm]{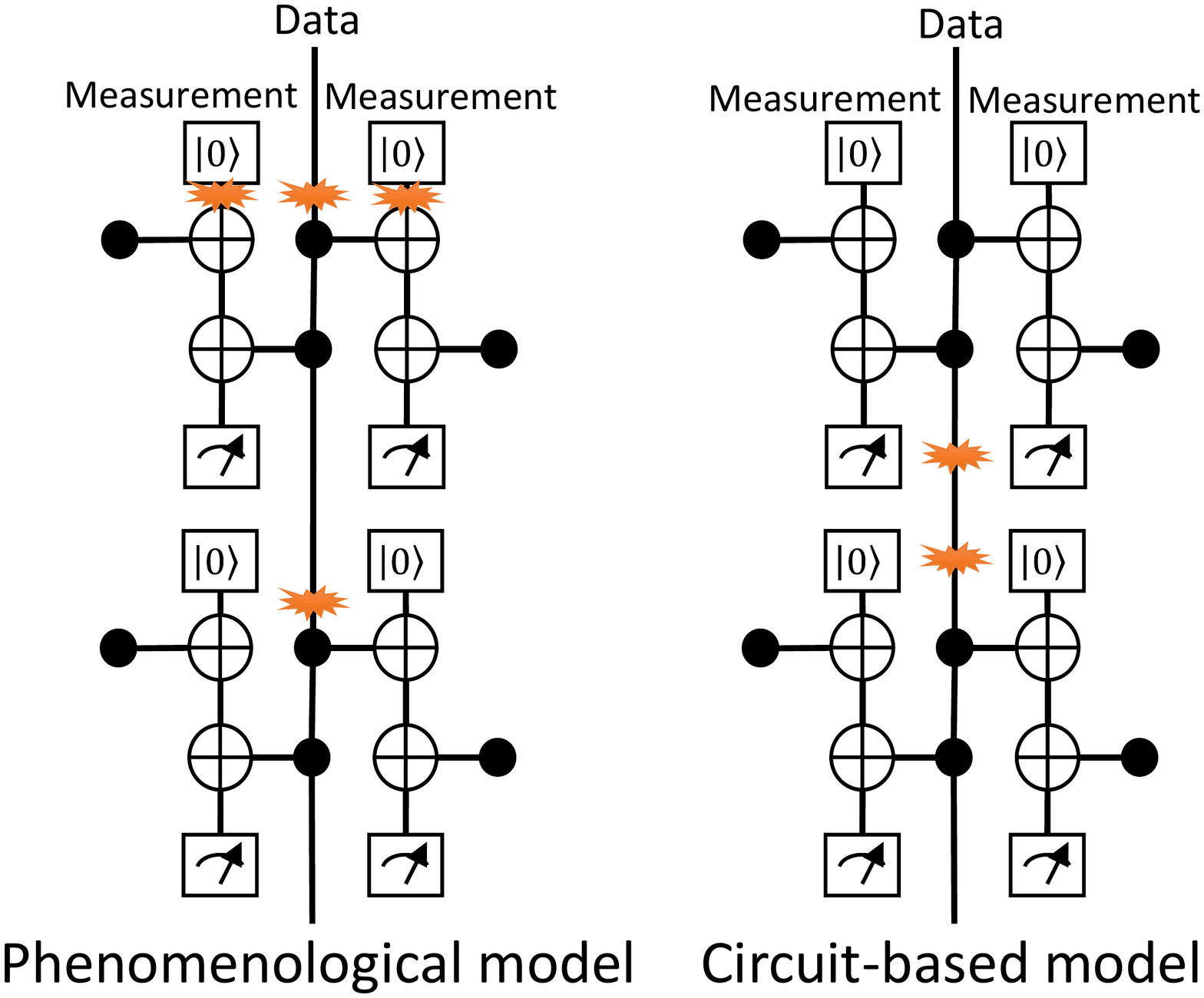}
 \caption{The sets of coherent error allocations which contribute to the probability distribution of the syndrome measurements.}
 \label{fig:set}
\end{figure}
Dependence of the error threshold $p_{\rm th}$ on the noise coherence $c$ is shown in the insets of Fig.\,\ref{fig:graph}. We see that the error threshold $p_{\rm th}$ decreases as the noise coherence $c$ increases. Note that non-uniform weighting improves the error threshold, but only slightly (see Appendix C \cite{SM}). The dependence on $c$ can be explained with a leading-order analysis as follows. 
The noise map of Eq.(\ref{eqnoise}) can be rewritten as ${\mathcal E}(\rho,X) = (1-p) \rho + ic\sqrt{(1-p)p}(X\rho-\rho X) + p X\rho X$. We call the second term $ic\sqrt{(1-p)p}(X \rho - \rho X)$ as the coherence term. This term contributes to diagonal terms of the density matrix  only through a concatenation of multiple noise maps. The correction to the diagonal terms after several cycles is written as even-order terms in $c\sqrt{(1-p)p}$. For $p \ll 1$, the leading order of the correction is $O(p)$ in the circuit-based model, while it is $O(p^2)$ in the phenomenological model since an error on a data qubit spreads to two measurement qubits before the next noise map is applied on the data qubit. For example, the product of the coherence terms of noise maps located in the positions shown in Fig.\,\ref{fig:set} contributes the correction.
In the case of the phenomenological model, the leading term is proportional to $c^4 p^2$, and its sign depends on the results of previous syndrome measurements. Such a noise leads to space-time correlations in the syndrome measurements. Since the decoder is not adapted to such correlations, the existence of coherence in noise is expected to result in a worse logical error probability.
On the other hand, in the case of the circuit-based model, the leading term is proportional to $c^2 p$, and it always increases the error probability. This directly worsens the logical error probability and the error threshold. 

In the case of the circuit-based model, we seek a more quantitative explanation of the behavior by proposing a heuristic ansatz as follows. We define an effective physical error probability of a data qubit per cycle $p_{\rm eff}(p,c)$  (the precise definition is given in Appendix D \cite{SM}). The probability $p_{\rm eff}(p,c)$ should be expanded for small $p$ as $p_{\rm eff}(p,c) = \beta(1 + \alpha c^2) p + O(p^2)$, where $\alpha$ is constant and is independent of the system size $n$. We assume that the logical error probability $p_{\rm L}(p,c)$ can be well explained by the local increase of noise, i.e., $p_{\rm L}(p,c) = p_{\rm L}((1+\alpha c^2)p , 0)$. Based on this ansatz, the error threshold under coherent noise $p_{\rm th}(c)$ can be written as $p_{\rm th}(c) \sim \frac{p_{\rm th}(0)}{1+\alpha c^2}$ if $p_{\rm th} \ll 1$. By using the analytically obtained value $\alpha=11/6$ (see Appendix D \cite{SM}), this ansatz gives the solid curve in the inset of Fig.\,\ref{fig:graph}, which is in good agreement with the accurate numerical results. We may expect that a similar leading-order ansatz also holds for the surface code, since it is a two-dimensional extension of the 1D repetition code. The factor $\alpha$ is also easily obtained by analytically calculating the effective bit-flip probability, and $p_{\rm th}(0)$ for incoherent noises can be efficiently computed. Therefore, the error threshold of the surface code under coherent noise will be estimated by the same approach.

{\it Conclusion and discussion.---}
We constructed an efficient and accurate scheme for estimating the error threshold of the 1D repetition code under coherent noise. 
We have calculated the error threshold under coherent noise in terms of the physical error probability $p$ and the noise coherence $c$. The parameters $p$ and $c$ can be experimentally accessible by randomized and purity benchmarkings, respectively \cite{knill2008randomized,joel2015estimating}. 
We emphasize here that the proposed accurate and efficient scheme is not limited to the 1D repetition code. In fact, in Appendix F \cite{SM} we provide an example with a fully quantum code, which simulates the surface code under a phenomenological coherent noise model. 
We have also proposed a leading-order ansatz for the estimation of the error threshold under coherent noise, and found that it reproduces the accurate numerical results well. This suggests that the effect of the coherent noise on the surface code will be assessed by an analogous ansatz, which can be calculated easily. In more general terms, the obtained accurate error thresholds of the 1D repetition code will serve as a reference to test the accuracy of approximation or heuristic schemes for simulating non-Clifford noise, as was done for the leading-order ansatz.
\begin{acknowledgments}
YS and KF thank Takanori Sugiyama for motivating an efficient simulation of QEC under coherent error, and for helpful discussions about quantum error characterization. This work is supported by KAKENHI No.16H02211, PRESTO, JST, CREST, JST and ERATO, JST. YS is supported by Advanced Leading Graduate Course for Photon Science.
\end{acknowledgments}

\pagebreak
\begin{widetext}
\begin{center}
\textbf{\large Efficient simulation of quantum error correction under coherent error based on non-unitary free-fermionic formalism - Supplemental material}
\end{center}
\end{widetext}
\setcounter{equation}{0}
\setcounter{figure}{0}
\setcounter{table}{0}
\setcounter{page}{1}
\makeatletter
\renewcommand{\theequation}{S\arabic{equation}}
\renewcommand{\thefigure}{S\arabic{figure}}

\appendix
\section{Appendix A: Detail of sampling scheme}
In this appendix, we describe the scheme of sampling $\bm{t}$ and computing $\Gamma_{\rm L}(t)/\Gamma(t)$ in detail. The simulation can be divided into three processes. 

{\it Process 1 --- Allocation of single-qubit noise maps \\}
In the case of the phenomenological model, the allocation of the noise maps is fixed and follows Fig.\,1(b). In the case of the circuit-based model, it is probabilistically chosen for each sampling as follows. The two-qubit noise map assumed after each CNOT gate contains a non-local $X$-type noise term $e^{iX_{\rm control} X_{\rm target}}$, which is not an FGO. We can convert this non-local noise to local noise by replacing it with a local noise preceding the CNOT gate $U_{\rm CNOT}$ since ${\mathcal E} (U_{\rm CNOT} \rho U_{\rm CNOT}^{\dagger} ,X_{\rm control} X_{\rm target})   = U_{\rm CNOT} {\mathcal E}(\rho,X_{\rm control} I_{\rm target}) U_{\rm CNOT}^{\dagger}$. Thus, to simulate the two-qubit noise map faithfully, we probabilistically place a single-qubit noise map ${\mathcal E}$ at 1) the target qubit after the CNOT gate, 2) the control qubit after the CNOT gate, and 3) the control qubit before the CNOT gate with the probabilities $p_{XI},p_{IX}$ and $p_{XX}$, respectively. The single-qubit noises associated with state preparations and measurements for the measurement qubits are placed deterministically. The ones for the data qubits at the beginning of the decoding process are also placed deterministically.

{\it Process 2 --- Simulation of the circuit\\}
The covariance matrix of the state $\ket{\tilde{\psi}} = (\ket{0}^{\otimes (n+1)}+\ket{1}^{\otimes (n+1)})/\sqrt{2}$ is given as follows.
\begin{eqnarray}
\tilde{M} &=& 
\left(\begin{array}{ccccccccc}
  0 & 0 & & & &&&& -1 \\ 
  0 & 0 & -1 & &&&&& \\
    & 1 & 0 & 0& &&&&\\
    &  & 0 & 0& -1&&&&\\
    &  &  & 1& 0&&&&\\
    &  &  & & &\ddots&&&\\
   & & & & &&0&-1& \\
   & & & & &&1&0&0 \\
  1 & & & & &&&0&0
\end{array} \right)
\end{eqnarray}
We start the simulation from $(\tilde{M} , 1)$, which is formally denoted by $(M({\bm t}_0), \Gamma({\bm t}_0))$. Given $(M({\bm t}_{k-1}),\Gamma({\bm t}_{k-1}))$, the value of $t_k$ is sampled from the probability $\Gamma({\bm t}_k)/\Gamma({\bm t}_{k-1})$.
Then the updated pair $(M({\bm t}_k),\Gamma({\bm t}_k))$ is calculated by 
\begin{widetext}
\begin{eqnarray}
M({\bm t}_k) &=& A^{(t_k)}-B^{(t_k)}(M({\bm t}_{k-1})-D^{(t_k)})^{-1} (B^{(t_k)})^T , \\
\Gamma({\bm t}_k) &=& \Gamma_{\rm G}^{(t_k)} \Gamma({\bm t}_{k-1}) \sqrt{\det(M({\bm t}_{k-1})-D^{(t_k)})} .
\end{eqnarray}
\end{widetext}
where $A^{(t_k)},B^{(t_k)},D^{(t_k)}$ and $\Gamma_{\rm G}^{(t_k)}$ are associated with the FGO $K_k^{(t_k)}$.

There are three types of operators $K_{{\rm noise},i}^{(\phi)}, K_{{\rm parity},i}^{(s,\phi)}$ and $K_{{\rm parity},i}^{(s)}$ for $K_k^{(t_k)}$. For $K_{{\rm noise},i}^{(\phi)} = \sqrt{p(\phi)}e^{i\phi X_i}$ with $p(\pm \theta) = \frac{1\pm c}{2}$, which is a noise operation on a data qubit, we have $\Gamma_{\rm G}^{(t_k)} = p(\phi)$ and $A^{(t_k)} = D^{(t_k)} = 0$. The submatrix $B^{(t_k)}$ is calculated as $B^{(t_k)} = I + B'$, where $B'$ has nonzero elements only for $\left(\begin{array}{cc} B'_{2i-1,2i-1} & B'_{2i-1,2i} \\ B'_{2i,2i-1} & B'_{2i,2i} \end{array}\right) = \left(\begin{array}{cc} -1+\cos 2\phi & -\sin 2 \phi \\ \sin 2 \phi & -1+\cos 2 \phi \end{array}\right)$.

For $K_{{\rm parity},i}^{(s,\phi)} = \sqrt{p(\phi)} \frac{1}{2} (I +(-1)^s e^{-2i\phi} Z_i Z_{i+1})$, which represents a parity measurement on two qubits, we have $\Gamma_{\rm G}^{(t_k)} = p(\phi)/2$. The matrix $A^{(t_k)}$ has nonzero elements only for $A_{2i,2i+1} = -A_{2i+1,2i} = -(-1)^s\cos 2 \phi$, and $D^{(t_k)} = -A^{(t_k)}$. The matrix $B^{(t_k)}$ is calculated as $B^{(t_k)} = I + B''$, where $B''$ has nonzero elements only for $\left(\begin{array}{cc} B''_{2i,2i} & B''_{2i,2i+1} \\ B''_{2i+1,2i} & B''_{2i+1,2i+1} \end{array}\right) = \left(\begin{array}{cc} -1 & (-1)^s \sin 2 \phi \\ -(-1)^s\sin 2 \phi & -1\end{array}\right)$.
The form of $p(\phi)$ depends on the noise allocation model. In the case of the phenomenological model, there is one $X$-type noise map on each measurement qubit, and the probability mass function $p(\phi)$ is defined as $p(\pm \theta) = \frac{1\pm c}{2}$. In the case of the circuit-based model, there are multiple $X$-type noise maps on the measurement qubit. We denote the number of the noise maps as $N$, which depends on the allocation at Process 1. Since a CNOT gate commutes with $X$-type noise map on the target qubit, we are allowed to use the same form of $K_{{\rm parity},i}^{(\phi)}$ except that the probability mass function $p(\phi)$ is replaced by $p(k\theta) = \binom {N} {(N-k)/2} \left(\frac{1+c}{2}\right)^{(N-k)/2} \left(\frac{1-c}{2}\right)^{(N+k)/2}$ for $k = \{-N, -N+2, ... , N-2, N \}$. 

Finally, for $K_{{\rm parity},i}^{(s)}$, we have $\Gamma_{\rm G}^{(t_k)}=\frac{1}{2}$, and $A^{(t_k)}, B^{(t_k)}$ and $D^{(t_k)}$ are equivalent to those for the $K_{{\rm parity},i}^{(s,\phi)}$ with $\phi=0$.

After repeating the $(n-1)T$ syndrome measurements, we perform noiseless parity measurement for each neighboring data qubits. As a result, we obtain $(n-1)(T+1)$ outputs of the parity measurements $\bm s$ included in $\bm t$, and the final state of data qubits $(M({\bm t}),\Gamma({\bm t})) = (M({\bm t}_{k_f}),\Gamma({\bm t}_{k_f}))$. 

{\it Process 3 --- Decoding \\}
We determine the recovery operation $R({\bm t})=\prod_{i=1}^{n} X_i^{r_i({\bm t})}$ by using minimum-weight perfect matching.  We then calculate $\Gamma_{\rm L}({\bm t})/\Gamma({\bm t})$ from Eqs. (5) and (6) as 
\begin{eqnarray}
\frac{\Gamma_{\rm L}({\bm t})}{\Gamma({\bm t})} &=& \frac{1 - (-1)^{r_n({\bm t})}M({\bm t})_{2n+1, 2n}} {2} .
\end{eqnarray}

Here we give a brief explanation of the minimum-weight perfect matching (see \cite{kelly2015state} for a detailed explanation). We denote the measurement outcome of the $x$-th measurement qubit at the $y$-th cycle as $s_{x,y}$, and denote $m_{x,y}=s_{x,y}\oplus s_{x,y-1}$ with $s_{x,0}=0$, and $m_{\rm bnd} = \oplus m_{x,y}$. The minimum-weight perfect matching can then be regarded as finding the most probable error pattern in the following empirical model associated with a weighted graph $G$ on the set of vertices $V=\{(x,y)\} \cup \{{\rm bnd}\}$: starting with $m_v=0$ for all $v \in V$, an error occurs on each edge $(v,v')$ independently with probability $e^{-w(v,v')}$, where $w(v,v')$ is the weight of $(v,v')$. Whenever an error occurs on an edge $(v,v')$, both $m_v$ and $m_{v'}$ are flipped.
For incoherent noises, the statistics of the actual circuit exactly follows the above model for an appropriate choice of $G$. In the case of the phenomenological model, $G$ is a uniformly weighted square lattice. For coherent noises, the statistics of the actual circuit does not exactly follow the simple model, and hence we need to choose $G$ heuristically. For simplicity, we choose the uniformly weighted square lattice for all the results shown in the main text. The possibility of using other graphs is discussed in Appendix C.

{\it Time efficiency\\}
Since there are $O(n^2)$ noise maps and syndrome measurements, and each of them takes at most $O(n^3)$ steps, this scheme requires $O(n^5)$ for each sampling. The time for simulation per sample with single thread of Intel Core i7 6700 takes about 20 $\rm ms$ with the parameter $(n,p,c)=(15,0.03,0)$ in the circuit-based model. 
For minimum-weight perfect matching, we used a library known as Kolmogorov's implementation of Edmonds' algorithm for minimum-weight perfect matching \cite{edmonds1965paths,kolmogorov2009blossom}.

\section{Appendix B: Obtaining the error threshold}
For each model and each value of $c$, the error threshold $p_{\rm th}$ was determined by the following procedure. For various values of physical error probability $p$ and code size $n$, the logical error probability $p_{\rm L}(p,n)$ was computed as an average over 50,000 samples. Then it was fitted to the function $p_{\rm L}(p,n) = a+b(p-p_{\rm th})n^{1/d}$ \cite{wang2003confinement,stephens2014fault} around the error threshold, where $a,b,d$ and $p_{\rm th}$ are fitting parameters. 

\section{Appendix C: Uniform and non-uniform weight in minimum-weight perfect matching}
\begin{figure}[tp]
   \centering
          \includegraphics[clip, width=7.5cm]{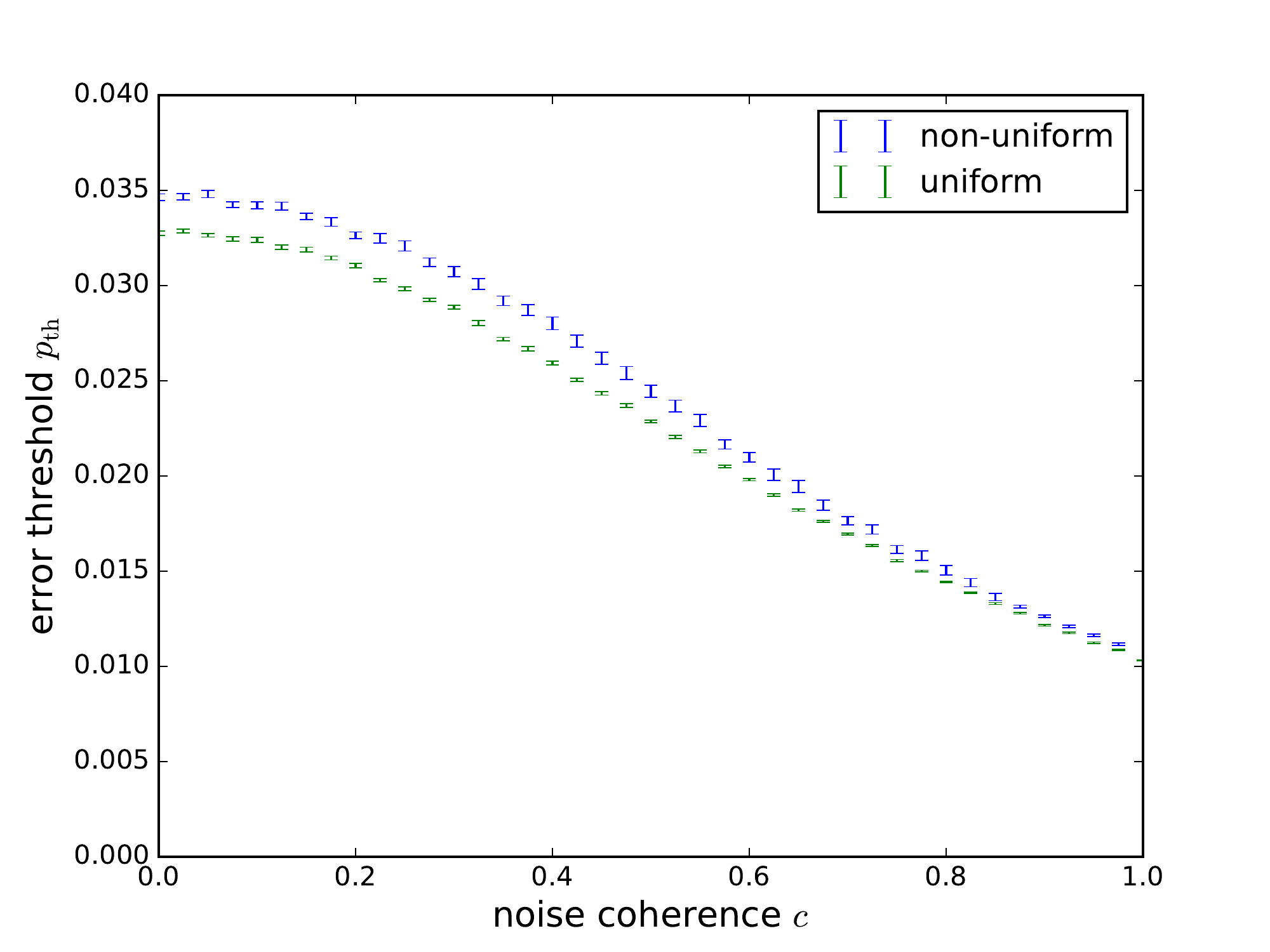}
          \caption{The error threshold $p_{\rm th}$ versus coherence $c$ in the circuit-based model. Two plots are correspond to uniformly weighted decoder and optimally weighted decoder.}
          \label{fig:weight_graph}
\end{figure}
As mentioned in Process 3 of Appendix A, it is possible to find a weighted graph that faithfully reproduces the statistics of the actual circuit if the noises are incoherent. In the case of the circuit-based model, such a faithful graph has non-uniform weights and diagonal edges \cite{kelly2015state}.

We constructed the faithful graph for the circuit-based model under incoherent noise, where each value of $e^{-w(v,v')}$ is approximated with its leading $O(p)$ term for simplicity. We applied the decoder based on the constructed graph to the circuit-based model with various values of coherence $c$. The obtained error thresholds are shown in Fig.\,\ref{fig:weight_graph}, together with the thresholds for the uniform-weight decoder. Compared to the error threshold using the uniform weight, the error threshold is improved for arbitrary values of $c$, but the amount of the improvement is small and the dependence of the error threshold on the noise coherence $c$ is also similar.

\section{Appendix D: Definition and calculation of the effective physical error probability}
We define an effective physical error probability of a data qubit per cycle $p_{\rm eff}(p,c)$ as a marginal probability with which results of two measurement qubits neighboring a data qubit are flipped at a certain cycle from the results of the previous cycle. More precisely, we define $p_{\rm eff}(p,c)$ as the marginal probability of $m_{x,y}=m_{x+1,y}=1$, using the notation introduced in Process 3 of Appendix A.
While $p_{\rm eff}(p,c)$ may depend on the values of $x$ and $y$, its leading term for small $p$ is independent of $x$ and $y$ (except $y=1$) and of the system size $n$. This leading term can be analytically obtained as $\frac{8}{3}(1+\frac{11}{6}c^2)p$, and thus $\alpha=11/6$. This definition can be simply generalized to the case of the surface code, and we can analytically obtain the effective physical error probability of a data qubit per cycle since only a few noise maps and qubits are relevant to the leading term.

\section{Appendix E: Logical error scaling below threshold}
\begin{figure*}[tp]
 \centering
  \subfloat[]{
 \includegraphics[clip, width=7.5cm]{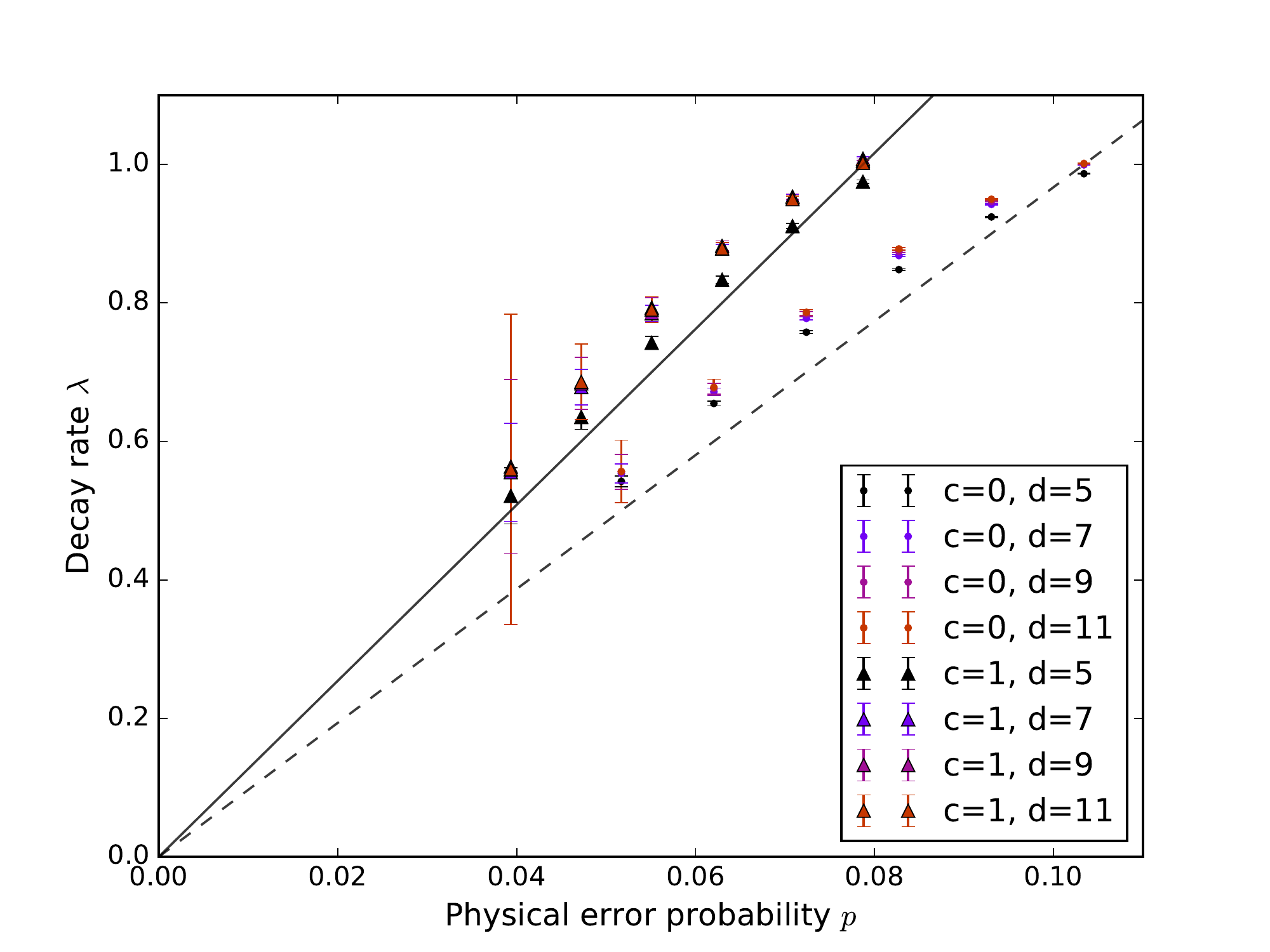}
  }
  \subfloat[]{
 \includegraphics[clip, width=7.5cm]{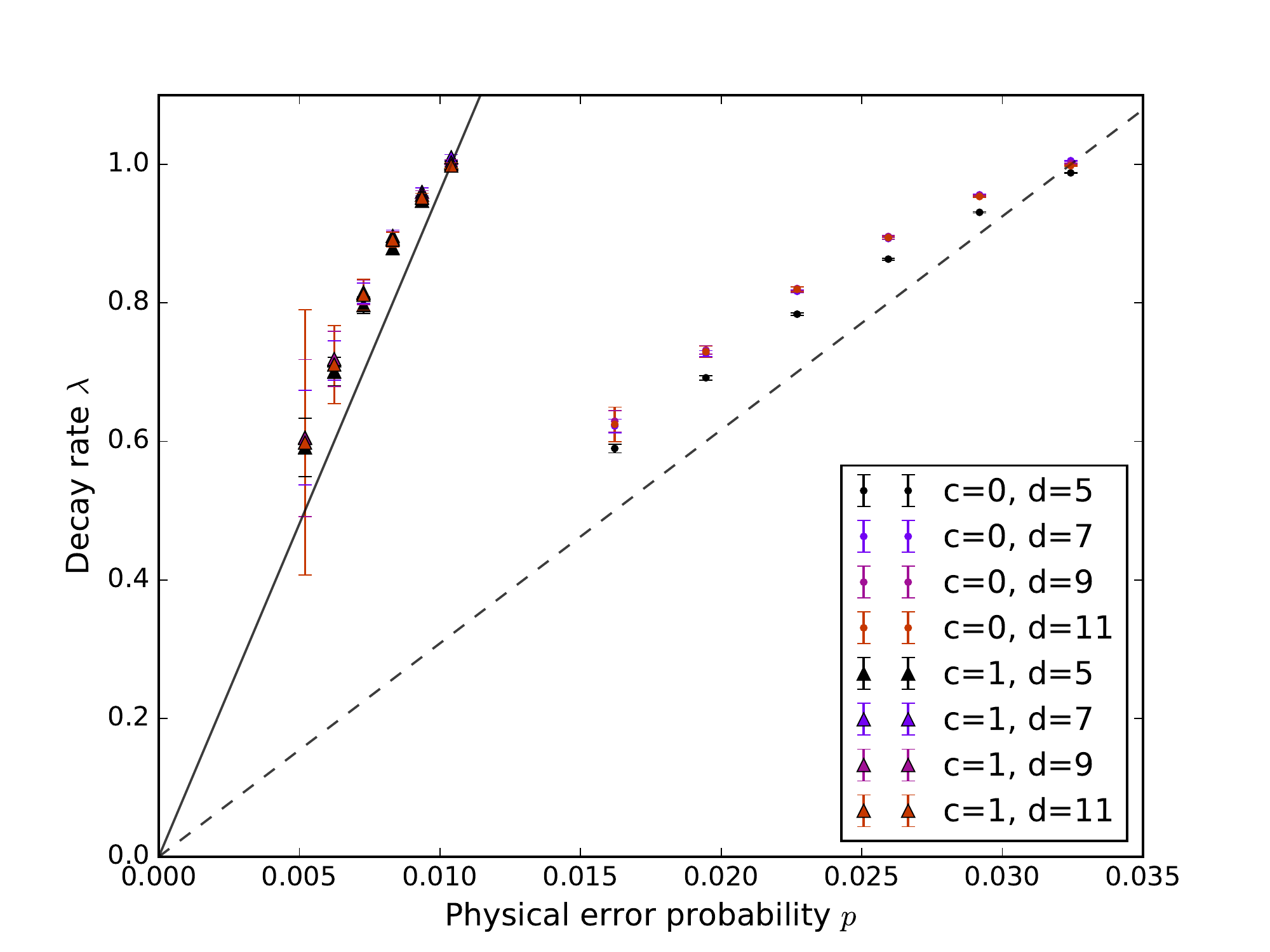}
  }
 \caption{The figures show the decay rate parameter $\lambda(d,p) := \frac{p_{\rm L} (d+2,p) }{ p_{\rm L}(d,p) }$ in terms of the physical error probability $p$, (a) for the phenomenological model and (b) for the circuit-based model. 
The cases with the incoherent noise ($c=0$) and those with the fully coherent noise ($c=1$) are marked with circles and triangles, respectively. The dashed and solid lines in each figure correspond to $\lambda=\frac{p}{p_{\rm th}}$ with $p_{\rm th}$ determined from the threshold behavior for $c=0$ and $1$, respectively.
The black symbols are for $d=5$, and the red ones are for $d=11$.}
 \label{fig:drop}
\end{figure*}

If the noise is incoherent and the physical error probability $p$ is below the threshold, the logical error probability is expected to decay with code distance $d$ as 
\begin{eqnarray}
\label{dropping}
p_{\rm L}(d,p) \sim \alpha \left( \frac{p}{p_{\rm th}}\right)^{d/2},
\end{eqnarray}
where $\alpha$ is a polynomial function of $d$ \cite{jones2012layered,raussendorf2007topological}. In this appendix, we numerically investigate whether this approximation is still valid for the coherent noises. Since $\alpha$ depends only weakly on $d$, the relation (\ref{dropping}) can be rewritten as 
\begin{eqnarray}
\lambda(d,p):= \frac{p_{\rm L}(d+2,p)}{p_{\rm L}(d,p)} \sim \frac{p}{p_{\rm th}},
\end{eqnarray}
We determined the above parameter $\lambda(d,p)$ from the computed values of $p_{\rm L}(d,p)$ and compared to the threshold $p_{\rm th}$ determined from the scaling ansatz. The result is shown in Fig.\,\ref{fig:drop} for the case of the coherent noise ($c = 1$) and the incoherent noise ($c = 0$).
In both the phenomenological and the circuit-based model, Eq.\,(\ref{dropping}) is satisfied with the same level of approximation regardless of the degree of coherence in the noises.

\section{Appendix F: The non-unitary free-fermionic formalism of the surface code}
Here we extend our scheme of the efficient classical simulation using non-unitary free-fermionic formalism to the surface code with a specific noise model including coherent errors.
We consider the surface code with $d^2$ data qubits and $(d^2-1)$ measurement qubits for an odd number $d$. An example with $d=5$ is shown in Fig.\,S3(a). 
\begin{figure*}[tp]
 \centering
  \subfloat[]{
 \includegraphics[clip ,width=10cm]{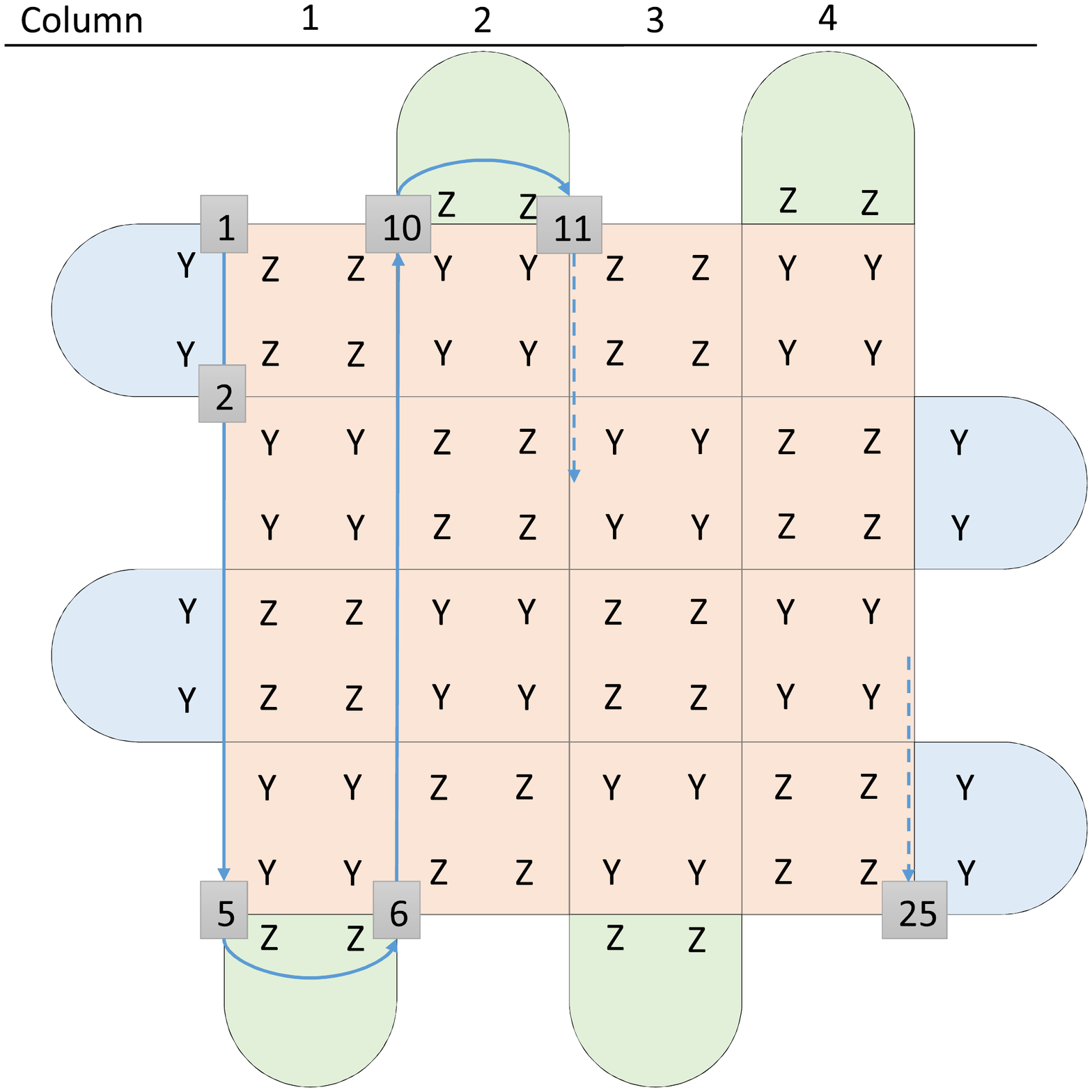}
 \label{fig:surface}
  }
  \subfloat[]{
 \includegraphics[clip ,width=5cm]{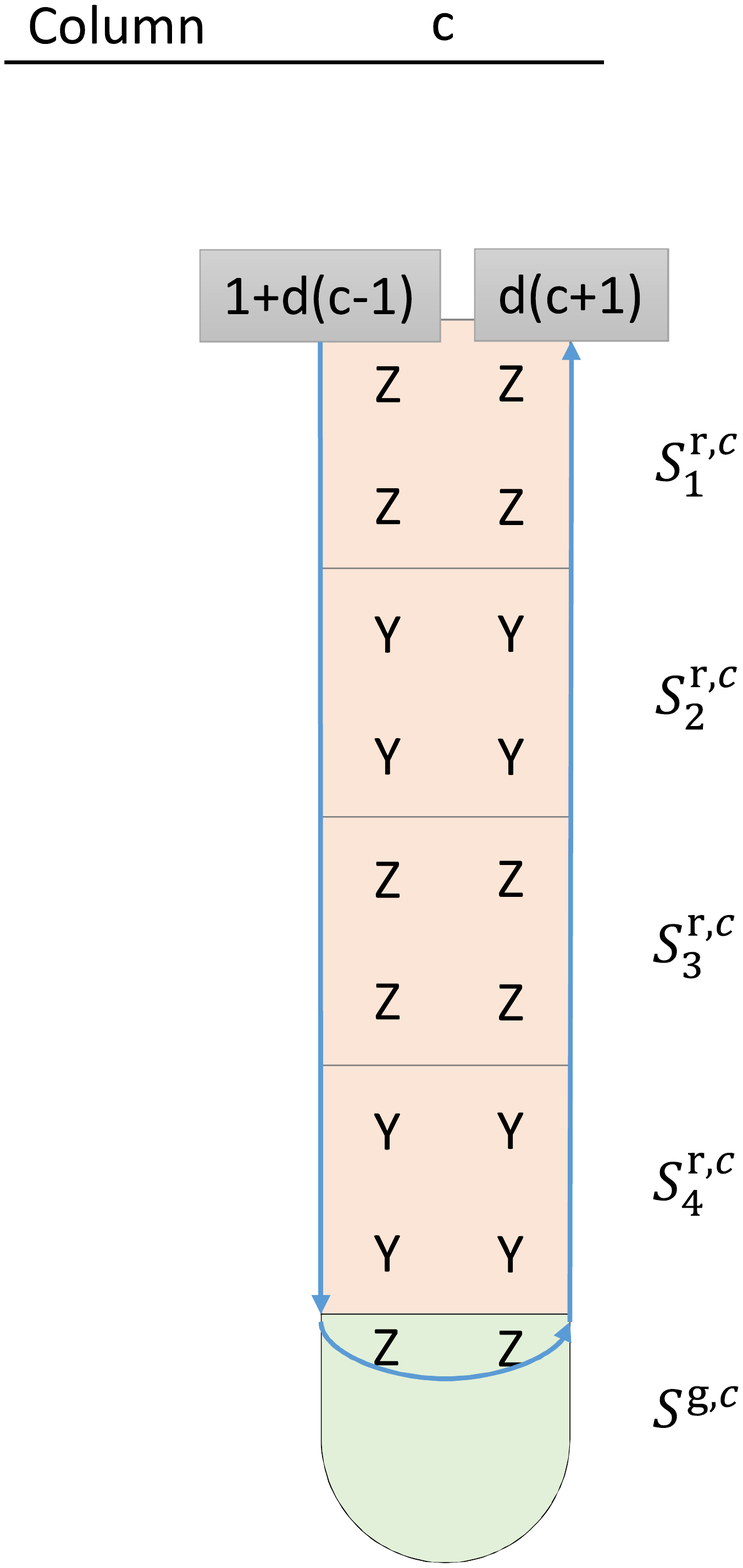}
 \label{fig:surface_part}
  }
 \caption{(a) The figure shows the qubit allocation of the surface code in the distance $d=5$. This architecture consists of $d^2$ data qubits and $d^2-1$ measurement qubits. See appendix F for the detail. (b) The figure shows a line of the surface code.}
\end{figure*}
The data qubits are located on the $d^2$ vertices of the colored faces. Each colored face corresponds to a measurement qubit. The measurement operator of each face is the product of $Y$ or $Z$ Pauli operators on the data qubits on its vertices. 
We denote the stabilizer operator of the blue face at the $r$-th row as $S_r^{\rm b}$ ($r=1,\ldots,d-1$). We denote the stabilizer operators of the red faces in the $c$-th column ($c = 1, \ldots, d-1$) as $\{S^{{\rm r},c}_i\}$ ($i=1,\ldots,d-1$), and that of the green face as $S^{{\rm g},c}$. The rules for assigning the index $i$ will be explained later. For each cycle of syndrome measurement, the stabilizer operators as observables are measured via controlled gates and $Z$-basis measurement on the measurement qubits.

We consider an error model in which the code truly works as a fully quantum code, namely, including both $X$- and $Z$-type errors. More specifically, we assume the following phenomenological noise model. 
For each cycle, one of the following three types of errors occurs on each data qubit probabilistically, the Pauli $Y$ error, the Pauli $Z$ error, and an $X$-type coherent error as in Eq.\,(1) in the main text.
We assume that the measurement qubits also suffer from the three types of errors probabilistically, the Pauli $X$, $Y$, and $Z$ errors. We also assume, for simplicity, that the syndrome measurement in the final cycle is error-free. This noise model can be considered as the phenomenological model in the main text with added Pauli $Y$ and $Z$ noises on both types of qubits, while limiting the $X$-type coherent errors to the data qubits.
Apparently, such an error model including the coherent noise cannot be treated with the method based on the Gottesman-Knill theorem. 

In contrast to the 1D repetition code in the main text where the logical error probability is the only parameter of interest, a circuit of fully quantum correction may be characterized in many different ways. Since we assumed that the final syndrome is correct, the whole circuit including a recovery operation can be viewed as a one-qubit channel $\mathcal{C}$ on the logical qubit space spanned by $\{\ket{0_{\rm L}},\ket{1_{\rm L}}\}$. The most general characterization of $\mathcal{C}$ is achieved if we learn the density matrix $\rho(\mathcal{C})$ on the logical qubit and an auxiliary qubit defined by
\begin{eqnarray}
\rho(\mathcal{C}) := (\mathcal{C} \otimes {\rm Id}) (\ket{\phi_{\rm init}} \bra{\phi_{\rm init}})
\end{eqnarray}
with 
\begin{eqnarray}
\ket{\phi_{\rm init}}:= \frac{\ket{0_{\rm L}} \ket{0} + \ket{1_{\rm L}} \ket{1}}{\sqrt{2}},
\end{eqnarray}
where ${\rm Id}$ is the identity channel. Any input-output relation can be calculated from $\rho(\mathcal{C})$, as well as the parameters such as various flip errors and fidelities.

Since the fermionic representation depends on the order of the data qubits, we assign numbers $1,2,\ldots,d^2 (=n)$ to the data qubits in the order as shown in the gray boxes in Fig.\,S3(a). 
We also assume that the ancillary qubit is the $(n+1)$-th data qubit. 
Recall that the Majorana fermionic operators can be chosen as
\begin{eqnarray}
c_{2i-1} &=& \left(\prod_{j=1}^{i-1}X_j\right) Z_i \\
c_{2i} &=& \left(\prod_{j=1}^{i-1}X_j \right) Y_i ,
\end{eqnarray}
where $i=1, \ldots, n+1$. 
While the stabilizer operators for the blue and green faces are quadratic, those for the red faces are not. 
Our idea is to introduce a new equivalent set of stabilizer operators which are all quadratic.
Let us consider the stabilizer operators in the $c$-th column, $\{S^{{\rm r},c}_1, \ldots S^{{\rm r},c}_{d-1}, S^{{\rm g},c} \}$. 
With an appropriate rotation, each column can be considered as a part shown in Fig.\,S3(b). We assign the index of the stabilizer operators of the red faces in the column as shown in the figure. 
Let us introduce new red stabilizer operators defined from $\{S^{{\rm r},c}_1, \ldots S^{{\rm r},c}_{d-1}, S^{{\rm g},c} \}$ as 
\begin{eqnarray}
\tilde{S}^{{\rm r},c}_i = (\prod_{j=i}^{d-1} {S}^{{\rm r},c}_j) S^{{\rm g},c}
\end{eqnarray}
for $i=1, \ldots , d-1$. Since they are explicitly written in the form
\begin{widetext}
\begin{eqnarray}
W_{d(c-1)+j} X_{d(c-1)+j+1} \cdots X_{d(c+1)-j} W_{d(c+1)+1-j},
\end{eqnarray}
\end{widetext}
where $W$ is either $Z$ or $Y$, they are all quadratic in Majorana fermionic operators.
The set $\{S^{{\rm r},c}_1, \ldots S^{{\rm r},c}_{d-1}, S^{{\rm g},c} \}$ can be conversely expressed in terms of $\{\tilde{S}^{{\rm r},c}_1, \ldots \tilde{S}^{{\rm r},c}_{d-1}, S^{{\rm g},c} \}$ as
\begin{eqnarray}
S^{{\rm r},c}_i = \begin{cases} \tilde{S}^{{\rm r},c}_i \tilde{S}^{{\rm r},c}_{i+1}  & \text{if } i<d-1 \\ \tilde{S}^{{\rm r},c}_{d-1} S^{{\rm g},c} & \text{if } i=d-1 \end{cases}.
\end{eqnarray}
This indicates that $\{S^{{\rm r},c}_1, \ldots S^{{\rm r},c}_{d-1}, S^{{\rm g},c} \}$ and $\{\tilde{S}^{{\rm r},c}_1, \ldots \tilde{S}^{{\rm r},c}_{d-1}, S^{{\rm g},c} \}$ are equivalent  sets as stabilizer generators. To be precise, let us introduce the projector onto the eigenspace of a stabilizer operator $S$ associated with syndrome bit $s$ by
\begin{eqnarray}
P(S,s) := \frac{1}{2}(I+(-1)^s S).
\end{eqnarray}
Then, it follows that
\begin{widetext}
\begin{eqnarray}
\label{syndeq}
P(S^{{\rm g},c},s^{\rm g}) \prod_{i=1}^{d-1} P(S^{{\rm g},c}_i,s^{\rm r}_i) = P(S^{{\rm g},c},s^{\rm g}) \prod_{i=1}^{d-1} P(\tilde{S}^{{\rm g},c}_i, \tilde{s}^{\rm r}_i )
\end{eqnarray}
\end{widetext}
if
\begin{eqnarray}
\label{syndeq2}
s_i= \begin{cases} (\tilde{s}^{\rm r}_i + \tilde{s}^{\rm r}_{i+1}) \mod 2 & \text{if } j<d-1 \\ (\tilde{s}^{\rm r}_{d-1} + s^{\rm g}) \mod 2 & \text{if } j=d-1  \end{cases}.
\end{eqnarray}
This implies that the syndrome measurement of $\{S^{{\rm r},c}_1, \ldots S^{{\rm r},c}_{d-1}, S^{{\rm g},c} \}$ can be equivalently done by those of $\{\tilde{S}^{{\rm r},c}_1, \ldots \tilde{S}^{{\rm r},c}_{d-1}, S^{{\rm g},c} \}$ followed by calculation according to Eq.\,(\ref{syndeq2}).

Now we will show that the quantum error correction circuit can be efficiently simulated to compute $\rho(\mathcal{C})$. First, we show that the initial state $\ket{\phi_{\rm init}}$ is an FGS.
The projection operator to the code space is given by 
\begin{eqnarray}
\mathcal{P}_{\rm C} &=& \left( \prod_{S \in \{S_i^{\rm b}\} \cup \{S^{{\rm r},c}_i\} \cup \{S^{{\rm g},c}\} } P(S,0) \right) \nonumber \\
&=& \left( \prod_{S \in \{S_i^{\rm b}\} \cup \{\tilde{S}^{{\rm r},c}_i\} \cup \{S^{{\rm g},c}\} } P(S,0) \right),
\end{eqnarray}
which is an FGO.
We consider the following logical $Z$ and $Y$ operators,
\begin{eqnarray}
L_Z &=& Z_{n-d+1} \left(\prod_{i=n-d+2}^{n} X_i \right) \label{LZ}, \\
L_Y &=& Y_{d} \left(\prod_{i=d+1}^{n} X_i \right) \label{LY},
\end{eqnarray}
and choose $\ket{0_{\rm L}}$ and $\ket{1_{\rm L}}$ to satisfy $L_Z \ket{0_{\rm L}} = \ket{0_{\rm L}}$, $L_Z \ket{1_{\rm L}} = -\ket{1_{\rm L}}$, and $L_Y \ket{0_{\rm L}} = i\ket{1_{\rm L}}$.
Since $\ket{\phi_{\rm init}}$ is stabilized by $L_Z Z_{n+1}$, $-L_YY_{n+1}$, and the $d^2-1$ stabilizer operators, we have 
\begin{widetext}
\begin{eqnarray}
\ket{\phi_{\rm init}}\bra{\phi_{\rm init}} = \mathcal{P}_{\rm C} P(L_ZZ_{n+1},0) P(-L_YY_{n+1},0).
\end{eqnarray}
\end{widetext}
Since $L_ZZ_{n+1}$ and $L_YY_{n+1}$ are quadratic, $\ket{\phi_{\rm init}}\bra{\phi_{\rm init}}$ is an FGO, and hence $\ket{\phi_{\rm init}}$ is an FGS.

Simulation of the execution of the circuit is done as follows. 
The error on each data qubit is sampled, and if it is $X$-type, the state can be directly updated to a new FGS since the $X$-type error operator is an FGO. Pauli $Z$ and $Y$ operator is not an FGO, and hence it cannot be directly applied. When we implement $Z$ error on the $i$-th qubit, we apply $Z_i Z_{n+1}$ instead, which is an FGO, and the updated state can be calculated. We apply $Y_i Z_{n+1}$ for $Y$ error. 
The appearance of $Z_{n+1}$ should be recorded, and will be compensated in the final stage. A syndrome measurement of a stabilizer operator $S$ on a current FGS $\ket{\phi}$ without any error is done by calculating 
\begin{eqnarray}
p_{s'} := \braket{\phi | P(S,s') | \phi} (s'=0,1),
\end{eqnarray}
and sampling $s'$ accordingly, and then calculating the updated state 
\begin{eqnarray}
\frac{P(S,s')\ket{\phi}} {\sqrt{\braket{\phi | P(S,s') | \phi}}}.
\end{eqnarray}
As we have seen, we are allowed to use $\tilde{S}^{{\rm r},c}_i$ instead of $S^{{\rm r},c}_i$, and to compute the syndrome for $\{S^{{\rm r},c}_i\}$ through Eq.\,(\ref{syndeq2}), which assures that $P(S,s')$ is always an FGO. 
As for the Pauli error on the measurement qubit, it can be equivalently translated to a composition of the following operations: a bit-flip of the computed syndrome (for $X$ and $Y$), and Pauli errors on the data qubit (for $Y$ and $Z$). 
After all the cycles are executed, the recovery operator $R$ is calculated using the sampled syndrome values. $R$ can be written as a product of Pauli operators and hence computed by possible inclusion of $Z_{n+1}$.

After a single run of simulation, we obtain a final FGS $\ket{\phi_{\rm final}}$ and the record of the number $w$ of applications of $Z_{n+1}$. If $w$ is even, the FGS is an accurate sample of the desired state $\rho(\mathcal{C})$. If $w$ is odd, $Z_{n+1}\ket{\phi_{\rm final}}$ is an accurate sample of $\rho(\mathcal{C})$. Hence, all we need is to convert the description of $\ket{\phi_{\rm final}}$ as an $(n+1)$-qubit FGS to that as a two-qubit density operator. 
We choose the Pauli operators for the logical qubit as $L_Z$ and $L_Y$ defined in Eqs. (\ref{LZ}) and (\ref{LY}), $L_I := I^{\otimes n}$, and 
\begin{eqnarray}
L_X = -i L_Y L_Z = - Y_d \left(\prod_{i=d+1}^{n-d}X_i \right) Y_{n-d+1}.
\end{eqnarray}
The density operator is then decomposed as 
\begin{eqnarray}
\ket{\phi_{\rm final}} \bra{\phi_{\rm final}} &=& \sum_{W,W'} \frac{1}{4} A_{W,W'} L_W W'_{n+1} \\
A_{W,W'} &:=& \braket{\phi_{\rm final} | L_W W'_{n+1} | \phi_{\rm final}},
\end{eqnarray}
where $W,W' \in \{I,X,Y,Z\}$ and $A_{I,I}=1$.
The coefficient $A_{X,X}$ can be obtained as follows. 
The final state $\ket{\phi_{\rm final}}$ is a +1 eigenstate of all the stabilizer operators. 
The operator $L_X$ can be transformed to $X^{\otimes n}$ by multiplying all the $(d-1)$ blue stabilizers, all the $(d-1)$ green stabilizers, and the $(d-1)^2/2$ red stabilizers composed of Pauli $Z$s. 
We thus have $L_X X_{n+1} \ket{\phi_{\rm final}}= X^{\otimes (n+1)} \ket{\phi_{\rm final}}$.
Since $X^{\otimes (n+1)}$ commutes with all the FGOs and $\ket{\phi_{\rm init}}$ is a +1 eigenstate of $X^{\otimes (n+1)}$, we have $X^{\otimes (n+1)} \ket{\phi_{\rm final}}=\ket{\phi_{\rm final}}$, and hence $A_{X,X}=1$.
This also implies $A_{W,W'}=0$ if $L_WW'_{n+1}$ anti-commutes with $L_X X_{n+1}$. The remaining 6 non-trivial coefficients are expectation values of $L_X, X_{n+1}, L_YY_{n+1}, L_YZ_{n+1}, L_ZY_{n+1}$, and $L_Z,Z_{n+1}$ which are all expectation values of FGOs. 
These values can be calculated from the FGS description of $\ket{\phi_{\rm final}}$. We thus obtain the density operator $\ket{\phi_{\rm final}} \bra{\phi_{\rm final}}$.

An accurate sample of $\rho(\mathcal{C})$ is then given by applying the correction $Z_{n+1}^w$.
The $\rho(\mathcal{C})$ is calculated as
\begin{eqnarray}
\rho(\mathcal{C}) = {\rm ave}(Z_{n+1}^w \ket{\phi_{\rm final}}\bra{\phi_{\rm final}} Z_{n+1}^w),
\end{eqnarray} 
where $\rm{ave}(\cdot)$ represents averaging function over the samples.

Various parameters associated with channel $\mathcal{C}$ can be calculated from $\rho(\mathcal{C})$. For example, the entanglement fidelity $F(\mathcal{C})$ represents how the channel preserves the input quantum state. This is computed as 
\begin{widetext}
\begin{eqnarray}
F(\mathcal{C}) &:=& \braket{\phi_{\rm init} | \rho(\mathcal{C}) | \phi_{\rm init}} \nonumber \\
&=& \frac{1}{4}( 1 + {\rm tr}(\rho(\mathcal{C})L_XX_{n+1}) - {\rm tr}(\rho(\mathcal{C})L_YY_{n+1})+{\rm tr}(\rho(\mathcal{C})L_ZZ_{n+1})) \nonumber \\
&=& \frac{1}{4} (1 + {\rm ave}( (-1)^w(1-A_{Y,Y}) + A_{Z,Z} )).
\end{eqnarray}
\end{widetext}

Finally, we would like to introduce another fully quantum error correcting circuit which is efficiently simulated by our method. The circuit is a modification of the previous circuit. It uses the same surface code, but each of the new red syndromes $\{ \tilde{S}^{{\rm r},c}_i \}$ is {\em actually} measured through a measurement qubit, instead of $\{ S^{{\rm r},c}_i \}$. For this circuit, we can allow $X$-type coherent errors on measurement qubits, just as in the phenomenological model in the main text.
In this case, the $X$-type coherent error can be absorbed by replacing the measurement operator $P(S,s)$ with 
\begin{eqnarray}
P'(S,s,\theta) := \frac{1}{2}(I+(-1)^s e^{-2i\theta} S),
\end{eqnarray}
where $\theta$ is a rotation angle dictated by the $X$-type coherent error. Since all the stabilizer operators are quadratic in the modified circuit, all of these operators are FGOs.

The second example shows that the 1D repetition code in the main text can be extended to a fully quantum code with no compromise on the noise model. Hence the applicability of our method relies neither on the simple structure of the repetition code nor on the absence of $Z$ and $Y$ errors. 
On the other hand, the comparison between the two examples of the surface code reveals an interesting trade-off. In the second example, the new syndrome measurements are non-local in the column direction, and only local in the row direction. In other words, it may be regarded as a 1D circuit with local stabilizer measurements. The first example is a true 2D circuit, but the efficient simulation seems to be possible only when the measurement qubits suffer no coherent errors.  
It is an open problem whether we can achieve both of them, efficient simulation of a truly 2D quantum error correction circuit with local syndrome measurement under coherent errors on both the data and measurement qubits.

\bibliographystyle{apsrev4-1}
\bibliography{cite_arxiv}

\begin{thebibliography}{48}%
\makeatletter
\providecommand \@ifxundefined [1]{%
 \@ifx{#1\undefined}
}%
\providecommand \@ifnum [1]{%
 \ifnum #1\expandafter \@firstoftwo
 \else \expandafter \@secondoftwo
 \fi
}%
\providecommand \@ifx [1]{%
 \ifx #1\expandafter \@firstoftwo
 \else \expandafter \@secondoftwo
 \fi
}%
\providecommand \natexlab [1]{#1}%
\providecommand \enquote  [1]{``#1''}%
\providecommand \bibnamefont  [1]{#1}%
\providecommand \bibfnamefont [1]{#1}%
\providecommand \citenamefont [1]{#1}%
\providecommand \href@noop [0]{\@secondoftwo}%
\providecommand \href [0]{\begingroup \@sanitize@url \@href}%
\providecommand \@href[1]{\@@startlink{#1}\@@href}%
\providecommand \@@href[1]{\endgroup#1\@@endlink}%
\providecommand \@sanitize@url [0]{\catcode `\\12\catcode `\$12\catcode
  `\&12\catcode `\#12\catcode `\^12\catcode `\_12\catcode `\%12\relax}%
\providecommand \@@startlink[1]{}%
\providecommand \@@endlink[0]{}%
\providecommand \url  [0]{\begingroup\@sanitize@url \@url }%
\providecommand \@url [1]{\endgroup\@href {#1}{\urlprefix }}%
\providecommand \urlprefix  [0]{URL }%
\providecommand \Eprint [0]{\href }%
\providecommand \doibase [0]{http://dx.doi.org/}%
\providecommand \selectlanguage [0]{\@gobble}%
\providecommand \bibinfo  [0]{\@secondoftwo}%
\providecommand \bibfield  [0]{\@secondoftwo}%
\providecommand \translation [1]{[#1]}%
\providecommand \BibitemOpen [0]{}%
\providecommand \bibitemStop [0]{}%
\providecommand \bibitemNoStop [0]{.\EOS\space}%
\providecommand \EOS [0]{\spacefactor3000\relax}%
\providecommand \BibitemShut  [1]{\csname bibitem#1\endcsname}%
\let\auto@bib@innerbib\@empty
\bibitem [{\citenamefont {Kitaev}(1997)}]{kitaev1997quantum}%
  \BibitemOpen
  \bibfield  {author} {\bibinfo {author} {\bibfnamefont {A.~Y.}\ \bibnamefont
  {Kitaev}},\ }\href@noop {} {\bibfield  {journal} {\bibinfo  {journal}
  {Russian Mathematical Surveys}\ }\textbf {\bibinfo {volume} {52}},\ \bibinfo
  {pages} {1191} (\bibinfo {year} {1997})}\BibitemShut {NoStop}%
\bibitem [{\citenamefont {Knill}\ \emph {et~al.}(1998)\citenamefont {Knill},
  \citenamefont {Laflamme},\ and\ \citenamefont {Zurek}}]{knill1998resilient}%
  \BibitemOpen
  \bibfield  {author} {\bibinfo {author} {\bibfnamefont {E.}~\bibnamefont
  {Knill}}, \bibinfo {author} {\bibfnamefont {R.}~\bibnamefont {Laflamme}}, \
  and\ \bibinfo {author} {\bibfnamefont {W.~H.}\ \bibnamefont {Zurek}},\ }in\
  \href@noop {} {\emph {\bibinfo {booktitle} {Proceedings of the Royal Society
  of London A: Mathematical, Physical and Engineering Sciences}}},\ Vol.\
  \bibinfo {volume} {454}\ (\bibinfo {organization} {The Royal Society},\
  \bibinfo {year} {1998})\ pp.\ \bibinfo {pages} {365--384}\BibitemShut
  {NoStop}%
\bibitem [{\citenamefont {Aharonov}\ and\ \citenamefont
  {Ben-Or}(1997)}]{aharonov1997fault}%
  \BibitemOpen
  \bibfield  {author} {\bibinfo {author} {\bibfnamefont {D.}~\bibnamefont
  {Aharonov}}\ and\ \bibinfo {author} {\bibfnamefont {M.}~\bibnamefont
  {Ben-Or}},\ }in\ \href@noop {} {\emph {\bibinfo {booktitle} {Proceedings of
  the twenty-ninth annual ACM symposium on Theory of computing}}}\ (\bibinfo
  {organization} {ACM},\ \bibinfo {year} {1997})\ pp.\ \bibinfo {pages}
  {176--188}\BibitemShut {NoStop}%
\bibitem [{\citenamefont {Fern}\ \emph {et~al.}(2006)\citenamefont {Fern},
  \citenamefont {Kempe}, \citenamefont {Simic},\ and\ \citenamefont
  {Sastry}}]{fern2004generalized}%
  \BibitemOpen
  \bibfield  {author} {\bibinfo {author} {\bibfnamefont {J.}~\bibnamefont
  {Fern}}, \bibinfo {author} {\bibfnamefont {J.}~\bibnamefont {Kempe}},
  \bibinfo {author} {\bibfnamefont {S.}~\bibnamefont {Simic}}, \ and\ \bibinfo
  {author} {\bibfnamefont {S.}~\bibnamefont {Sastry}},\ }in\ \href@noop {}
  {\emph {\bibinfo {booktitle} {IEEE Trans. on Automatic Control}}},\
  Vol.~\bibinfo {volume} {51}\ (\bibinfo {year} {2006})\ pp.\ \bibinfo {pages}
  {448--459}\BibitemShut {NoStop}%
\bibitem [{\citenamefont {Greenbaum}\ and\ \citenamefont
  {Dutton}(2016)}]{greenbaum2016modeling}%
  \BibitemOpen
  \bibfield  {author} {\bibinfo {author} {\bibfnamefont {D.}~\bibnamefont
  {Greenbaum}}\ and\ \bibinfo {author} {\bibfnamefont {Z.}~\bibnamefont
  {Dutton}},\ }\href@noop {} {\bibfield  {journal} {\bibinfo  {journal} {arXiv
  preprint arXiv:1612.03908}\ } (\bibinfo {year} {2016})}\BibitemShut {NoStop}%
\bibitem [{\citenamefont {Wang}\ \emph {et~al.}(2003)\citenamefont {Wang},
  \citenamefont {Harrington},\ and\ \citenamefont
  {Preskill}}]{wang2003confinement}%
  \BibitemOpen
  \bibfield  {author} {\bibinfo {author} {\bibfnamefont {C.}~\bibnamefont
  {Wang}}, \bibinfo {author} {\bibfnamefont {J.}~\bibnamefont {Harrington}}, \
  and\ \bibinfo {author} {\bibfnamefont {J.}~\bibnamefont {Preskill}},\
  }\href@noop {} {\bibfield  {journal} {\bibinfo  {journal} {Annals of
  Physics}\ }\textbf {\bibinfo {volume} {303}},\ \bibinfo {pages} {31}
  (\bibinfo {year} {2003})}\BibitemShut {NoStop}%
\bibitem [{\citenamefont {Wang}\ \emph {et~al.}(2011)\citenamefont {Wang},
  \citenamefont {Fowler},\ and\ \citenamefont {Hollenberg}}]{wang2011surface}%
  \BibitemOpen
  \bibfield  {author} {\bibinfo {author} {\bibfnamefont {D.~S.}\ \bibnamefont
  {Wang}}, \bibinfo {author} {\bibfnamefont {A.~G.}\ \bibnamefont {Fowler}}, \
  and\ \bibinfo {author} {\bibfnamefont {L.~C.~L.}\ \bibnamefont
  {Hollenberg}},\ }\href@noop {} {\bibfield  {journal} {\bibinfo  {journal}
  {Physical Review A}\ }\textbf {\bibinfo {volume} {83}},\ \bibinfo {pages}
  {020302} (\bibinfo {year} {2011})}\BibitemShut {NoStop}%
\bibitem [{\citenamefont {Fowler}\ \emph
  {et~al.}(2012{\natexlab{a}})\citenamefont {Fowler}, \citenamefont
  {Whiteside},\ and\ \citenamefont {Hollenberg}}]{fowler2012towards}%
  \BibitemOpen
  \bibfield  {author} {\bibinfo {author} {\bibfnamefont {A.~G.}\ \bibnamefont
  {Fowler}}, \bibinfo {author} {\bibfnamefont {A.~C.}\ \bibnamefont
  {Whiteside}}, \ and\ \bibinfo {author} {\bibfnamefont {L.~C.~L.}\
  \bibnamefont {Hollenberg}},\ }\href@noop {} {\bibfield  {journal} {\bibinfo
  {journal} {Phys. Rev. Lett.}\ }\textbf {\bibinfo {volume} {108}},\ \bibinfo
  {pages} {180501} (\bibinfo {year} {2012}{\natexlab{a}})}\BibitemShut
  {NoStop}%
\bibitem [{\citenamefont {Stephens}(2014)}]{stephens2014fault}%
  \BibitemOpen
  \bibfield  {author} {\bibinfo {author} {\bibfnamefont {A.~M.}\ \bibnamefont
  {Stephens}},\ }\href@noop {} {\bibfield  {journal} {\bibinfo  {journal}
  {Physical Review A}\ }\textbf {\bibinfo {volume} {89}},\ \bibinfo {pages}
  {022321} (\bibinfo {year} {2014})}\BibitemShut {NoStop}%
\bibitem [{\citenamefont {Ghosh}\ \emph {et~al.}(2012)\citenamefont {Ghosh},
  \citenamefont {Fowler},\ and\ \citenamefont {Geller}}]{ghosh2012surface}%
  \BibitemOpen
  \bibfield  {author} {\bibinfo {author} {\bibfnamefont {J.}~\bibnamefont
  {Ghosh}}, \bibinfo {author} {\bibfnamefont {A.~G.}\ \bibnamefont {Fowler}}, \
  and\ \bibinfo {author} {\bibfnamefont {M.~R.}\ \bibnamefont {Geller}},\
  }\href@noop {} {\bibfield  {journal} {\bibinfo  {journal} {Physical Review
  A}\ }\textbf {\bibinfo {volume} {86}},\ \bibinfo {pages} {062318} (\bibinfo
  {year} {2012})}\BibitemShut {NoStop}%
\bibitem [{\citenamefont {Geller}\ and\ \citenamefont
  {Zhou}(2013)}]{geller2013efficient}%
  \BibitemOpen
  \bibfield  {author} {\bibinfo {author} {\bibfnamefont {M.~R.}\ \bibnamefont
  {Geller}}\ and\ \bibinfo {author} {\bibfnamefont {Z.}~\bibnamefont {Zhou}},\
  }\href@noop {} {\bibfield  {journal} {\bibinfo  {journal} {Physical Review
  A}\ }\textbf {\bibinfo {volume} {88}},\ \bibinfo {pages} {012314} (\bibinfo
  {year} {2013})}\BibitemShut {NoStop}%
\bibitem [{\citenamefont {Guti{\'e}rrez}\ \emph {et~al.}(2013)\citenamefont
  {Guti{\'e}rrez}, \citenamefont {Svec}, \citenamefont {Vargo},\ and\
  \citenamefont {Brown}}]{gutierrez2013approximation}%
  \BibitemOpen
  \bibfield  {author} {\bibinfo {author} {\bibfnamefont {M.}~\bibnamefont
  {Guti{\'e}rrez}}, \bibinfo {author} {\bibfnamefont {L.}~\bibnamefont {Svec}},
  \bibinfo {author} {\bibfnamefont {A.}~\bibnamefont {Vargo}}, \ and\ \bibinfo
  {author} {\bibfnamefont {K.~R.}\ \bibnamefont {Brown}},\ }\href@noop {}
  {\bibfield  {journal} {\bibinfo  {journal} {Physical Review A}\ }\textbf
  {\bibinfo {volume} {87}},\ \bibinfo {pages} {030302} (\bibinfo {year}
  {2013})}\BibitemShut {NoStop}%
\bibitem [{\citenamefont {Guti{\'e}rrez}\ and\ \citenamefont
  {Brown}(2015)}]{gutierrez2015comparison}%
  \BibitemOpen
  \bibfield  {author} {\bibinfo {author} {\bibfnamefont {M.}~\bibnamefont
  {Guti{\'e}rrez}}\ and\ \bibinfo {author} {\bibfnamefont {K.~R.}\ \bibnamefont
  {Brown}},\ }\href@noop {} {\bibfield  {journal} {\bibinfo  {journal}
  {Physical Review A}\ }\textbf {\bibinfo {volume} {91}},\ \bibinfo {pages}
  {022335} (\bibinfo {year} {2015})}\BibitemShut {NoStop}%
\bibitem [{\citenamefont {Magesan}\ \emph {et~al.}(2013)\citenamefont
  {Magesan}, \citenamefont {Puzzuoli}, \citenamefont {Granade},\ and\
  \citenamefont {Cory}}]{magesan2013modeling}%
  \BibitemOpen
  \bibfield  {author} {\bibinfo {author} {\bibfnamefont {E.}~\bibnamefont
  {Magesan}}, \bibinfo {author} {\bibfnamefont {D.}~\bibnamefont {Puzzuoli}},
  \bibinfo {author} {\bibfnamefont {C.~E.}\ \bibnamefont {Granade}}, \ and\
  \bibinfo {author} {\bibfnamefont {D.~G.}\ \bibnamefont {Cory}},\ }\href@noop
  {} {\bibfield  {journal} {\bibinfo  {journal} {Physical Review A}\ }\textbf
  {\bibinfo {volume} {87}},\ \bibinfo {pages} {012324} (\bibinfo {year}
  {2013})}\BibitemShut {NoStop}%
\bibitem [{\citenamefont {Puzzuoli}\ \emph {et~al.}(2014)\citenamefont
  {Puzzuoli}, \citenamefont {Granade}, \citenamefont {Haas}, \citenamefont
  {Criger}, \citenamefont {Magesan},\ and\ \citenamefont
  {Cory}}]{puzzuoli2014tractable}%
  \BibitemOpen
  \bibfield  {author} {\bibinfo {author} {\bibfnamefont {D.}~\bibnamefont
  {Puzzuoli}}, \bibinfo {author} {\bibfnamefont {C.}~\bibnamefont {Granade}},
  \bibinfo {author} {\bibfnamefont {H.}~\bibnamefont {Haas}}, \bibinfo {author}
  {\bibfnamefont {B.}~\bibnamefont {Criger}}, \bibinfo {author} {\bibfnamefont
  {E.}~\bibnamefont {Magesan}}, \ and\ \bibinfo {author} {\bibfnamefont
  {D.~G.}\ \bibnamefont {Cory}},\ }\href@noop {} {\bibfield  {journal}
  {\bibinfo  {journal} {Physical Review A}\ }\textbf {\bibinfo {volume} {89}},\
  \bibinfo {pages} {022306} (\bibinfo {year} {2014})}\BibitemShut {NoStop}%
\bibitem [{\citenamefont {Guti{\'e}rrez}\ \emph {et~al.}(2016)\citenamefont
  {Guti{\'e}rrez}, \citenamefont {Smith}, \citenamefont {Lulushi},
  \citenamefont {Janardan},\ and\ \citenamefont {Brown}}]{gutierrez2016errors}%
  \BibitemOpen
  \bibfield  {author} {\bibinfo {author} {\bibfnamefont {M.}~\bibnamefont
  {Guti{\'e}rrez}}, \bibinfo {author} {\bibfnamefont {C.}~\bibnamefont
  {Smith}}, \bibinfo {author} {\bibfnamefont {L.}~\bibnamefont {Lulushi}},
  \bibinfo {author} {\bibfnamefont {S.}~\bibnamefont {Janardan}}, \ and\
  \bibinfo {author} {\bibfnamefont {K.~R.}\ \bibnamefont {Brown}},\ }\href@noop
  {} {\bibfield  {journal} {\bibinfo  {journal} {Physical Review A}\ }\textbf
  {\bibinfo {volume} {94}},\ \bibinfo {pages} {042338} (\bibinfo {year}
  {2016})}\BibitemShut {NoStop}%
\bibitem [{\citenamefont {Rahn}\ \emph {et~al.}(2002)\citenamefont {Rahn},
  \citenamefont {Doherty},\ and\ \citenamefont {Mabuchi}}]{rahn2002exact}%
  \BibitemOpen
  \bibfield  {author} {\bibinfo {author} {\bibfnamefont {B.}~\bibnamefont
  {Rahn}}, \bibinfo {author} {\bibfnamefont {A.~C.}\ \bibnamefont {Doherty}}, \
  and\ \bibinfo {author} {\bibfnamefont {H.}~\bibnamefont {Mabuchi}},\
  }\href@noop {} {\bibfield  {journal} {\bibinfo  {journal} {Physical Review
  A}\ }\textbf {\bibinfo {volume} {66}},\ \bibinfo {pages} {032304} (\bibinfo
  {year} {2002})}\BibitemShut {NoStop}%
\bibitem [{\citenamefont {Chamberland}\ \emph {et~al.}(2016)\citenamefont
  {Chamberland}, \citenamefont {Wallman}, \citenamefont {Beale},\ and\
  \citenamefont {Laflamme}}]{chamberland2016hard}%
  \BibitemOpen
  \bibfield  {author} {\bibinfo {author} {\bibfnamefont {C.}~\bibnamefont
  {Chamberland}}, \bibinfo {author} {\bibfnamefont {J.~J.}\ \bibnamefont
  {Wallman}}, \bibinfo {author} {\bibfnamefont {S.}~\bibnamefont {Beale}}, \
  and\ \bibinfo {author} {\bibfnamefont {R.}~\bibnamefont {Laflamme}},\
  }\href@noop {} {\bibfield  {journal} {\bibinfo  {journal} {arXiv preprint
  arXiv:1612.02830}\ } (\bibinfo {year} {2016})}\BibitemShut {NoStop}%
\bibitem [{\citenamefont {Tomita}\ and\ \citenamefont
  {Svore}(2014)}]{tomita2014low}%
  \BibitemOpen
  \bibfield  {author} {\bibinfo {author} {\bibfnamefont {Y.}~\bibnamefont
  {Tomita}}\ and\ \bibinfo {author} {\bibfnamefont {K.~M.}\ \bibnamefont
  {Svore}},\ }\href@noop {} {\bibfield  {journal} {\bibinfo  {journal}
  {Physical Review A}\ }\textbf {\bibinfo {volume} {90}},\ \bibinfo {pages}
  {062320} (\bibinfo {year} {2014})}\BibitemShut {NoStop}%
\bibitem [{\citenamefont {Ferris}\ and\ \citenamefont
  {Poulin}(2014)}]{ferris2014tensor}%
  \BibitemOpen
  \bibfield  {author} {\bibinfo {author} {\bibfnamefont {A.~J.}\ \bibnamefont
  {Ferris}}\ and\ \bibinfo {author} {\bibfnamefont {D.}~\bibnamefont
  {Poulin}},\ }\href@noop {} {\bibfield  {journal} {\bibinfo  {journal} {Phys.
  Rev. Lett.}\ }\textbf {\bibinfo {volume} {113}},\ \bibinfo {pages} {030501}
  (\bibinfo {year} {2014})}\BibitemShut {NoStop}%
\bibitem [{\citenamefont {Darmawan}\ and\ \citenamefont
  {Poulin}(2016)}]{darmawan2016tensor}%
  \BibitemOpen
  \bibfield  {author} {\bibinfo {author} {\bibfnamefont {A.~S.}\ \bibnamefont
  {Darmawan}}\ and\ \bibinfo {author} {\bibfnamefont {D.}~\bibnamefont
  {Poulin}},\ }\href@noop {} {\bibfield  {journal} {\bibinfo  {journal} {arXiv
  preprint arXiv:1607.06460}\ } (\bibinfo {year} {2016})}\BibitemShut {NoStop}%
\bibitem [{\citenamefont {Gottesman}(1998)}]{gottesman1998heisenberg}%
  \BibitemOpen
  \bibfield  {author} {\bibinfo {author} {\bibfnamefont {D.}~\bibnamefont
  {Gottesman}},\ }\href@noop {} {\bibfield  {journal} {\bibinfo  {journal}
  {arXiv preprint quant-ph/9807006}\ } (\bibinfo {year} {1998})}\BibitemShut
  {NoStop}%
\bibitem [{\citenamefont {Aaronson}\ and\ \citenamefont
  {Gottesman}(2004)}]{aaronson2004improved}%
  \BibitemOpen
  \bibfield  {author} {\bibinfo {author} {\bibfnamefont {S.}~\bibnamefont
  {Aaronson}}\ and\ \bibinfo {author} {\bibfnamefont {D.}~\bibnamefont
  {Gottesman}},\ }\href@noop {} {\bibfield  {journal} {\bibinfo  {journal}
  {Physical Review A}\ }\textbf {\bibinfo {volume} {70}},\ \bibinfo {pages}
  {052328} (\bibinfo {year} {2004})}\BibitemShut {NoStop}%
\bibitem [{\citenamefont {Kelly}\ \emph {et~al.}(2015)\citenamefont {Kelly},
  \citenamefont {Barends}, \citenamefont {Fowler}, \citenamefont {Megrant},
  \citenamefont {Jeffrey}, \citenamefont {White}, \citenamefont {Sank},
  \citenamefont {Mutus}, \citenamefont {Campbell}, \citenamefont {Chen} \emph
  {et~al.}}]{kelly2015state}%
  \BibitemOpen
  \bibfield  {author} {\bibinfo {author} {\bibfnamefont {J.}~\bibnamefont
  {Kelly}}, \bibinfo {author} {\bibfnamefont {R.}~\bibnamefont {Barends}},
  \bibinfo {author} {\bibfnamefont {A.}~\bibnamefont {Fowler}}, \bibinfo
  {author} {\bibfnamefont {A.}~\bibnamefont {Megrant}}, \bibinfo {author}
  {\bibfnamefont {E.}~\bibnamefont {Jeffrey}}, \bibinfo {author} {\bibfnamefont
  {T.}~\bibnamefont {White}}, \bibinfo {author} {\bibfnamefont
  {D.}~\bibnamefont {Sank}}, \bibinfo {author} {\bibfnamefont {J.}~\bibnamefont
  {Mutus}}, \bibinfo {author} {\bibfnamefont {B.}~\bibnamefont {Campbell}},
  \bibinfo {author} {\bibfnamefont {Y.}~\bibnamefont {Chen}},  \emph {et~al.},\
  }\href@noop {} {\bibfield  {journal} {\bibinfo  {journal} {Nature}\ }\textbf
  {\bibinfo {volume} {519}},\ \bibinfo {pages} {66} (\bibinfo {year}
  {2015})}\BibitemShut {NoStop}%
\bibitem [{\citenamefont {C{\'o}rcoles}\ \emph {et~al.}(2015)\citenamefont
  {C{\'o}rcoles}, \citenamefont {Magesan}, \citenamefont {Srinivasan},
  \citenamefont {Cross}, \citenamefont {Steffen}, \citenamefont {Gambetta},\
  and\ \citenamefont {Chow}}]{corcoles2015demonstration}%
  \BibitemOpen
  \bibfield  {author} {\bibinfo {author} {\bibfnamefont {A.}~\bibnamefont
  {C{\'o}rcoles}}, \bibinfo {author} {\bibfnamefont {E.}~\bibnamefont
  {Magesan}}, \bibinfo {author} {\bibfnamefont {S.~J.}\ \bibnamefont
  {Srinivasan}}, \bibinfo {author} {\bibfnamefont {A.~W.}\ \bibnamefont
  {Cross}}, \bibinfo {author} {\bibfnamefont {M.}~\bibnamefont {Steffen}},
  \bibinfo {author} {\bibfnamefont {J.~M.}\ \bibnamefont {Gambetta}}, \ and\
  \bibinfo {author} {\bibfnamefont {J.~M.}\ \bibnamefont {Chow}},\ }\href@noop
  {} {\bibfield  {journal} {\bibinfo  {journal} {Nature communications}\
  }\textbf {\bibinfo {volume} {6}} (\bibinfo {year} {2015})}\BibitemShut
  {NoStop}%
\bibitem [{\citenamefont {Rist{\`e}}\ \emph {et~al.}(2015)\citenamefont
  {Rist{\`e}}, \citenamefont {Poletto}, \citenamefont {Huang}, \citenamefont
  {Bruno}, \citenamefont {Vesterinen}, \citenamefont {Saira},\ and\
  \citenamefont {DiCarlo}}]{riste2015detecting}%
  \BibitemOpen
  \bibfield  {author} {\bibinfo {author} {\bibfnamefont {D.}~\bibnamefont
  {Rist{\`e}}}, \bibinfo {author} {\bibfnamefont {S.}~\bibnamefont {Poletto}},
  \bibinfo {author} {\bibfnamefont {M.-Z.}\ \bibnamefont {Huang}}, \bibinfo
  {author} {\bibfnamefont {A.}~\bibnamefont {Bruno}}, \bibinfo {author}
  {\bibfnamefont {V.}~\bibnamefont {Vesterinen}}, \bibinfo {author}
  {\bibfnamefont {O.-P.}\ \bibnamefont {Saira}}, \ and\ \bibinfo {author}
  {\bibfnamefont {L.}~\bibnamefont {DiCarlo}},\ }\href@noop {} {\bibfield
  {journal} {\bibinfo  {journal} {Nature communications}\ }\textbf {\bibinfo
  {volume} {6}} (\bibinfo {year} {2015})}\BibitemShut {NoStop}%
\bibitem [{\citenamefont {Kueng}\ \emph {et~al.}(2016)\citenamefont {Kueng},
  \citenamefont {Long}, \citenamefont {Doherty},\ and\ \citenamefont
  {Flammia}}]{kueng2015comparing}%
  \BibitemOpen
  \bibfield  {author} {\bibinfo {author} {\bibfnamefont {R.}~\bibnamefont
  {Kueng}}, \bibinfo {author} {\bibfnamefont {D.~M.}\ \bibnamefont {Long}},
  \bibinfo {author} {\bibfnamefont {A.~C.}\ \bibnamefont {Doherty}}, \ and\
  \bibinfo {author} {\bibfnamefont {S.~T.}\ \bibnamefont {Flammia}},\
  }\href@noop {} {\bibfield  {journal} {\bibinfo  {journal} {Phys. Rev. Lett.}\
  }\textbf {\bibinfo {volume} {117}},\ \bibinfo {pages} {170502} (\bibinfo
  {year} {2016})}\BibitemShut {NoStop}%
\bibitem [{\citenamefont {Lidar}\ and\ \citenamefont
  {Brun}(2013)}]{lidar2013quantum}%
  \BibitemOpen
  \bibfield  {author} {\bibinfo {author} {\bibfnamefont {D.~A.}\ \bibnamefont
  {Lidar}}\ and\ \bibinfo {author} {\bibfnamefont {T.~A.}\ \bibnamefont
  {Brun}},\ }\href@noop {} {\emph {\bibinfo {title} {Quantum error
  correction}}}\ (\bibinfo  {publisher} {Cambridge University Press},\ \bibinfo
  {year} {2013})\BibitemShut {NoStop}%
\bibitem [{\citenamefont {Valiant}(2002)}]{valiant2002quantum}%
  \BibitemOpen
  \bibfield  {author} {\bibinfo {author} {\bibfnamefont {L.~G.}\ \bibnamefont
  {Valiant}},\ }\href@noop {} {\bibfield  {journal} {\bibinfo  {journal} {SIAM
  Journal on Computing}\ }\textbf {\bibinfo {volume} {31}},\ \bibinfo {pages}
  {1229} (\bibinfo {year} {2002})}\BibitemShut {NoStop}%
\bibitem [{\citenamefont {Terhal}\ and\ \citenamefont
  {DiVincenzo}(2002)}]{terhal2002classical}%
  \BibitemOpen
  \bibfield  {author} {\bibinfo {author} {\bibfnamefont {B.~M.}\ \bibnamefont
  {Terhal}}\ and\ \bibinfo {author} {\bibfnamefont {D.~P.}\ \bibnamefont
  {DiVincenzo}},\ }\href@noop {} {\bibfield  {journal} {\bibinfo  {journal}
  {Physical Review A}\ }\textbf {\bibinfo {volume} {65}},\ \bibinfo {pages}
  {032325} (\bibinfo {year} {2002})}\BibitemShut {NoStop}%
\bibitem [{\citenamefont {Knill}(2001)}]{knill2001fermionic}%
  \BibitemOpen
  \bibfield  {author} {\bibinfo {author} {\bibfnamefont {E.}~\bibnamefont
  {Knill}},\ }\href@noop {} {\bibfield  {journal} {\bibinfo  {journal} {arXiv
  preprint quant-ph/0108033}\ } (\bibinfo {year} {2001})}\BibitemShut {NoStop}%
\bibitem [{\citenamefont {Bravyi}(2005{\natexlab{a}})}]{bravyi2004lagrangian}%
  \BibitemOpen
  \bibfield  {author} {\bibinfo {author} {\bibfnamefont {S.}~\bibnamefont
  {Bravyi}},\ }in\ \href@noop {} {\emph {\bibinfo {booktitle} {Quantum Inf. and
  Comp.}}},\ Vol.~\bibinfo {volume} {5}\ (\bibinfo {year} {2005})\ pp.\
  \bibinfo {pages} {216--238}\BibitemShut {NoStop}%
\bibitem [{\citenamefont {Bravyi}(2005{\natexlab{b}})}]{bravyi2005classical}%
  \BibitemOpen
  \bibfield  {author} {\bibinfo {author} {\bibfnamefont {S.}~\bibnamefont
  {Bravyi}},\ }\href@noop {} {\bibfield  {journal} {\bibinfo  {journal} {arXiv
  preprint quant-ph/0507282}\ } (\bibinfo {year}
  {2005}{\natexlab{b}})}\BibitemShut {NoStop}%
\bibitem [{\citenamefont {Jozsa}\ and\ \citenamefont
  {Miyake}(2008)}]{jozsa2008matchgates}%
  \BibitemOpen
  \bibfield  {author} {\bibinfo {author} {\bibfnamefont {R.}~\bibnamefont
  {Jozsa}}\ and\ \bibinfo {author} {\bibfnamefont {A.}~\bibnamefont {Miyake}},\
  }in\ \href@noop {} {\emph {\bibinfo {booktitle} {Proceedings of the Royal
  Society of London A: Mathematical, Physical and Engineering Sciences}}},\
  Vol.\ \bibinfo {volume} {464}\ (\bibinfo {organization} {The Royal Society},\
  \bibinfo {year} {2008})\ pp.\ \bibinfo {pages} {3089--3106}\BibitemShut
  {NoStop}%
\bibitem [{\citenamefont {Jozsa}\ \emph {et~al.}(2015)\citenamefont {Jozsa},
  \citenamefont {Miyake},\ and\ \citenamefont {Strelchuk}}]{jozsa2013jordan}%
  \BibitemOpen
  \bibfield  {author} {\bibinfo {author} {\bibfnamefont {R.}~\bibnamefont
  {Jozsa}}, \bibinfo {author} {\bibfnamefont {A.}~\bibnamefont {Miyake}}, \
  and\ \bibinfo {author} {\bibfnamefont {S.}~\bibnamefont {Strelchuk}},\ }in\
  \href@noop {} {\emph {\bibinfo {booktitle} {Quantum Inf. and Comp.}}},\
  Vol.~\bibinfo {volume} {15}\ (\bibinfo {year} {2015})\ pp.\ \bibinfo {pages}
  {541--556}\BibitemShut {NoStop}%
\bibitem [{\citenamefont {Brod}(2016)}]{brod2016efficient}%
  \BibitemOpen
  \bibfield  {author} {\bibinfo {author} {\bibfnamefont {D.~J.}\ \bibnamefont
  {Brod}},\ }\href@noop {} {\bibfield  {journal} {\bibinfo  {journal} {Physical
  Review A}\ }\textbf {\bibinfo {volume} {93}},\ \bibinfo {pages} {062332}
  (\bibinfo {year} {2016})}\BibitemShut {NoStop}%
\bibitem [{\citenamefont {Bravyi}\ \emph {et~al.}(2014)\citenamefont {Bravyi},
  \citenamefont {Suchara},\ and\ \citenamefont {Vargo}}]{bravyi2014efficient}%
  \BibitemOpen
  \bibfield  {author} {\bibinfo {author} {\bibfnamefont {S.}~\bibnamefont
  {Bravyi}}, \bibinfo {author} {\bibfnamefont {M.}~\bibnamefont {Suchara}}, \
  and\ \bibinfo {author} {\bibfnamefont {A.}~\bibnamefont {Vargo}},\
  }\href@noop {} {\bibfield  {journal} {\bibinfo  {journal} {Physical Review
  A}\ }\textbf {\bibinfo {volume} {90}},\ \bibinfo {pages} {032326} (\bibinfo
  {year} {2014})}\BibitemShut {NoStop}%
\bibitem [{\citenamefont {Dennis}\ \emph {et~al.}(2002)\citenamefont {Dennis},
  \citenamefont {Kitaev}, \citenamefont {Landahl},\ and\ \citenamefont
  {Preskill}}]{dennis2002topological}%
  \BibitemOpen
  \bibfield  {author} {\bibinfo {author} {\bibfnamefont {E.}~\bibnamefont
  {Dennis}}, \bibinfo {author} {\bibfnamefont {A.}~\bibnamefont {Kitaev}},
  \bibinfo {author} {\bibfnamefont {A.}~\bibnamefont {Landahl}}, \ and\
  \bibinfo {author} {\bibfnamefont {J.}~\bibnamefont {Preskill}},\ }\href@noop
  {} {\bibfield  {journal} {\bibinfo  {journal} {Journal of Mathematical
  Physics}\ }\textbf {\bibinfo {volume} {43}},\ \bibinfo {pages} {4452}
  (\bibinfo {year} {2002})}\BibitemShut {NoStop}%
\bibitem [{\citenamefont {Bravyi}\ and\ \citenamefont
  {Kitaev}(1998)}]{bravyi1998quantum}%
  \BibitemOpen
  \bibfield  {author} {\bibinfo {author} {\bibfnamefont {S.~B.}\ \bibnamefont
  {Bravyi}}\ and\ \bibinfo {author} {\bibfnamefont {A.~Y.}\ \bibnamefont
  {Kitaev}},\ }\href@noop {} {\bibfield  {journal} {\bibinfo  {journal} {arXiv
  preprint quant-ph/9811052}\ } (\bibinfo {year} {1998})}\BibitemShut {NoStop}%
\bibitem [{\citenamefont {Fowler}\ \emph
  {et~al.}(2012{\natexlab{b}})\citenamefont {Fowler}, \citenamefont
  {Mariantoni}, \citenamefont {Martinis},\ and\ \citenamefont
  {Cleland}}]{fowler2012surface}%
  \BibitemOpen
  \bibfield  {author} {\bibinfo {author} {\bibfnamefont {A.~G.}\ \bibnamefont
  {Fowler}}, \bibinfo {author} {\bibfnamefont {M.}~\bibnamefont {Mariantoni}},
  \bibinfo {author} {\bibfnamefont {J.~M.}\ \bibnamefont {Martinis}}, \ and\
  \bibinfo {author} {\bibfnamefont {A.~N.}\ \bibnamefont {Cleland}},\
  }\href@noop {} {\bibfield  {journal} {\bibinfo  {journal} {Physical Review
  A}\ }\textbf {\bibinfo {volume} {86}},\ \bibinfo {pages} {032324} (\bibinfo
  {year} {2012}{\natexlab{b}})}\BibitemShut {NoStop}%
\bibitem [{\citenamefont {Landahl}\ \emph {et~al.}(2011)\citenamefont
  {Landahl}, \citenamefont {Anderson},\ and\ \citenamefont
  {Rice}}]{landahl2011fault}%
  \BibitemOpen
  \bibfield  {author} {\bibinfo {author} {\bibfnamefont {A.~J.}\ \bibnamefont
  {Landahl}}, \bibinfo {author} {\bibfnamefont {J.~T.}\ \bibnamefont
  {Anderson}}, \ and\ \bibinfo {author} {\bibfnamefont {P.~R.}\ \bibnamefont
  {Rice}},\ }\href@noop {} {\bibfield  {journal} {\bibinfo  {journal} {arXiv
  preprint arXiv:1108.5738}\ } (\bibinfo {year} {2011})}\BibitemShut {NoStop}%
\bibitem [{SM()}]{SM}%
  \BibitemOpen
  \href@noop {} {}\bibinfo {note} {See supplemental material for details of
  sampling scheme, fitting method of the error threshold, details of
  minimum-weight perfect matching, definition of the effective physical error
  probability, dropping rate of the logical error probability after the
  threshold, and the method how to simulate the surface coder under coherent
  errors efficiently.}\BibitemShut {Stop}%
\bibitem [{\citenamefont {Edmonds}(1965)}]{edmonds1965paths}%
  \BibitemOpen
  \bibfield  {author} {\bibinfo {author} {\bibfnamefont {J.}~\bibnamefont
  {Edmonds}},\ }\href@noop {} {\bibfield  {journal} {\bibinfo  {journal}
  {Canadian Journal of mathematics}\ }\textbf {\bibinfo {volume} {17}},\
  \bibinfo {pages} {449} (\bibinfo {year} {1965})}\BibitemShut {NoStop}%
\bibitem [{\citenamefont {Kolmogorov}(2009)}]{kolmogorov2009blossom}%
  \BibitemOpen
  \bibfield  {author} {\bibinfo {author} {\bibfnamefont {V.}~\bibnamefont
  {Kolmogorov}},\ }\href@noop {} {\bibfield  {journal} {\bibinfo  {journal}
  {Mathematical Programming Computation}\ }\textbf {\bibinfo {volume} {1}},\
  \bibinfo {pages} {43} (\bibinfo {year} {2009})}\BibitemShut {NoStop}%
\bibitem [{\citenamefont {Knill}\ \emph {et~al.}(2008)\citenamefont {Knill},
  \citenamefont {Leibfried}, \citenamefont {Reichle}, \citenamefont {Britton},
  \citenamefont {Blakestad}, \citenamefont {Jost}, \citenamefont {Langer},
  \citenamefont {Ozeri}, \citenamefont {Seidelin},\ and\ \citenamefont
  {Wineland}}]{knill2008randomized}%
  \BibitemOpen
  \bibfield  {author} {\bibinfo {author} {\bibfnamefont {E.}~\bibnamefont
  {Knill}}, \bibinfo {author} {\bibfnamefont {D.}~\bibnamefont {Leibfried}},
  \bibinfo {author} {\bibfnamefont {R.}~\bibnamefont {Reichle}}, \bibinfo
  {author} {\bibfnamefont {J.}~\bibnamefont {Britton}}, \bibinfo {author}
  {\bibfnamefont {R.~B.}\ \bibnamefont {Blakestad}}, \bibinfo {author}
  {\bibfnamefont {J.~D.}\ \bibnamefont {Jost}}, \bibinfo {author}
  {\bibfnamefont {C.}~\bibnamefont {Langer}}, \bibinfo {author} {\bibfnamefont
  {R.}~\bibnamefont {Ozeri}}, \bibinfo {author} {\bibfnamefont
  {S.}~\bibnamefont {Seidelin}}, \ and\ \bibinfo {author} {\bibfnamefont
  {D.~J.}\ \bibnamefont {Wineland}},\ }\href@noop {} {\bibfield  {journal}
  {\bibinfo  {journal} {Phys. Rev. A}\ }\textbf {\bibinfo {volume} {77}},\
  \bibinfo {pages} {012307} (\bibinfo {year} {2008})}\BibitemShut {NoStop}%
\bibitem [{\citenamefont {Wallman}\ \emph {et~al.}(2015)\citenamefont
  {Wallman}, \citenamefont {Granade}, \citenamefont {Harper},\ and\
  \citenamefont {Flammia}}]{joel2015estimating}%
  \BibitemOpen
  \bibfield  {author} {\bibinfo {author} {\bibfnamefont {J.}~\bibnamefont
  {Wallman}}, \bibinfo {author} {\bibfnamefont {C.}~\bibnamefont {Granade}},
  \bibinfo {author} {\bibfnamefont {R.}~\bibnamefont {Harper}}, \ and\ \bibinfo
  {author} {\bibfnamefont {S.~T.}\ \bibnamefont {Flammia}},\ }\href@noop {}
  {\bibfield  {journal} {\bibinfo  {journal} {New Journal of Physics}\ }\textbf
  {\bibinfo {volume} {17}},\ \bibinfo {pages} {113020} (\bibinfo {year}
  {2015})}\BibitemShut {NoStop}%
\bibitem [{\citenamefont {Jones}\ \emph {et~al.}(2012)\citenamefont {Jones},
  \citenamefont {Van~Meter}, \citenamefont {Fowler}, \citenamefont {McMahon},
  \citenamefont {Kim}, \citenamefont {Ladd},\ and\ \citenamefont
  {Yamamoto}}]{jones2012layered}%
  \BibitemOpen
  \bibfield  {author} {\bibinfo {author} {\bibfnamefont {N.~C.}\ \bibnamefont
  {Jones}}, \bibinfo {author} {\bibfnamefont {R.}~\bibnamefont {Van~Meter}},
  \bibinfo {author} {\bibfnamefont {A.~G.}\ \bibnamefont {Fowler}}, \bibinfo
  {author} {\bibfnamefont {P.~L.}\ \bibnamefont {McMahon}}, \bibinfo {author}
  {\bibfnamefont {J.}~\bibnamefont {Kim}}, \bibinfo {author} {\bibfnamefont
  {T.~D.}\ \bibnamefont {Ladd}}, \ and\ \bibinfo {author} {\bibfnamefont
  {Y.}~\bibnamefont {Yamamoto}},\ }\href@noop {} {\bibfield  {journal}
  {\bibinfo  {journal} {Physical Review X}\ }\textbf {\bibinfo {volume} {2}},\
  \bibinfo {pages} {031007} (\bibinfo {year} {2012})}\BibitemShut {NoStop}%
\bibitem [{\citenamefont {Raussendorf}\ \emph {et~al.}(2007)\citenamefont
  {Raussendorf}, \citenamefont {Harrington},\ and\ \citenamefont
  {Goyal}}]{raussendorf2007topological}%
  \BibitemOpen
  \bibfield  {author} {\bibinfo {author} {\bibfnamefont {R.}~\bibnamefont
  {Raussendorf}}, \bibinfo {author} {\bibfnamefont {J.}~\bibnamefont
  {Harrington}}, \ and\ \bibinfo {author} {\bibfnamefont {K.}~\bibnamefont
  {Goyal}},\ }\href@noop {} {\bibfield  {journal} {\bibinfo  {journal} {New
  Journal of Physics}\ }\textbf {\bibinfo {volume} {9}},\ \bibinfo {pages}
  {199} (\bibinfo {year} {2007})}\BibitemShut {NoStop}%
\end{thebibliography}%

\end{document}